\documentclass[a4paper,fleqn,usenatbib,useAMS]{mnras}

\usepackage[T1]{fontenc}
\usepackage{ae,aecompl}
\usepackage{natbib}

\usepackage{graphicx}	
\usepackage{amsmath}	
\usepackage{amssymb}	
\usepackage{multicol}        
\usepackage{bm}		
\usepackage{pdflscape}	
\usepackage{booktabs}

\title[Rest-UV to z$\sim$5]{The Redshift Evolution of Rest-UV Spectroscopic Properties to z$\sim$5}

\author[Pahl et al.]{Anthony J. Pahl,$^{1}$\thanks{Contact e-mail: \href{mailto:pahl@astro.ucla.edu}{pahl@astro.ucla.edu}}
Alice Shapley,$^{1}$
Andreas L. Faisst,$^{2}$
Peter L. Capak,$^{2}$
Xinnan Du,$^{3}$\newauthor
Naveen A. Reddy,$^{3}$
Peter Laursen$^{4,5}$
and Michael W. Topping$^{1}$	
	\\
	$^{1}$Department of Physics and Astronomy, University of California, Los Angeles, CA 90095, USA\\
	$^{2}$IPAC, California Institute of Technology, 1200 East California Boulevard, Pasadena, CA 91125, USA\\
	$^{3}$Department of Physics and Astronomy, University of California Riverside, Riverside, CA 92521, USA\\
	$^{4}$Institute of Theoretical Astrophysics, University of Oslo, PO Box 1029 Blindern 0315 Oslo, Norway\\
	$^{5}$Cosmic Dawn Center (DAWN); Niels Bohr Institute, University of Copenhagen, Vibenshuset, Lyngbyvej 2, DK-2100 Copenhagen {\O}, Denmark}

\date{}

\pubyear{2019}

\begin{document}
	\label{firstpage}
	\pagerange{\pageref{firstpage}--\pageref{lastpage}}
	\maketitle

\begin{abstract}

We perform a comprehensive analysis of the redshift evolution of the rest-UV spectra of star-forming galaxies out to $z\sim5$. 
We combine new $z\sim5$ measurements of HI Ly$\alpha$ and low- and high-ionization interstellar metal absorption features with comparable measurements at $z\sim2\textrm{--}4$. 
We measure the equivalent widths of interstellar absorption features using stacked spectra in bins of Ly$\alpha$ equivalent width, performing corrections to Ly$\alpha$ strengths based on a model for the transmission of the intergalactic medium.
We find a strong correlation between decreasing low-ionization absorption strength and increasing Ly$\alpha$ emission strength over the redshift range $z\sim2\textrm{--}5$, suggesting that both of these quantities are fundamentally linked to neutral gas covering fraction.
At the highest Ly$\alpha$ equivalent widths, we observe evolution at $z\sim5$ towards greater Ly$\alpha$ emission strength at fixed low-ionization absorption strength. 
If we interpret the non-evolving relationship of Ly$\alpha$ emission strength and low-ionization line strength at $z\sim2-4$ as primarily reflecting the radiative transfer of Ly$\alpha$ photons, this evolution at $z\sim5$ suggests a higher intrinsic production rate of Ly$\alpha$ photons than at lower redshift.
Our conclusion is supported by the joint evolution of the relationships among Ly$\alpha$ emission strength, interstellar absorption strength, and dust reddening.
We perform additional analysis in bins of stellar mass, star-formation rate, UV luminosity, and age, examining how the relationships between galaxy properties and Ly$\alpha$ emission evolve towards higher redshift.
We conclude that increasing intrinsic Ly$\alpha$ photon production and strong detection of nebular C~\textsc{iv} emission (signaling lower metallicity) at $z\sim5$ indicate an elevated ionized photon production efficiency ($\xi_{\rm ion}$).

\end{abstract}

\begin{keywords}
galaxies: evolution -- galaxies: high-redshift -- galaxies: ISM
\end{keywords}

\section{Introduction} \label{sec:intro}

The rest-UV spectrum of a star-forming galaxy
provides a uniquely rich probe of the properties of massive stars and the interstellar and circumgalactic medium \citep[ISM and CGM;][]{Shapley2003,Steidel2010,Steidel2016,Leitherer2011,Senchyna2019}.
Understanding the interplay between these components is important for placing constraints on galaxy evolution across cosmic time. Furthermore,
the gas in and around distant galaxies that is probed by rest-UV spectra also modulates the escape of ionization
radiation. This gas provides important clues to the process of cosmic reionization, a key phase
transition in which the intergalactic medium (IGM) transformed from neutral to ionized by roughly
a billion years after the Big Bang at $z\sim6$ \citep{Fan2006,Robertson2015a,Planck2016}.

Measuring the rest-UV spectra of distant galaxies requires optical instrumentation and long
integrations. Based on observations with instruments such as the Low Resolution Imager and Spectrometer (LRIS) on
the Keck I telescope \citep{Oke1995,Steidel2004}, there have been several analyses of the relationships between rest-UV spectral features and other galaxy properties out to $z\sim4$ \citep[e.g.,][]{Du2018,Jones2012,Marchi2019a}. \citet{Du2018} (hereafter, D18) performed a comprehensive evolutionary analysis of rest-UV spectral trends of star forming galaxies from $z\sim2\textrm{--}4$. This analysis centred around the hydrogen Ly$\alpha$ line and
low- and high-ionization interstellar (LIS and HIS) metal absorption lines, which together probe the neutral gas covering fraction, gas kinematics, and ISM and CGM properties of galaxies. Both D18 and \citet{Jones2012} find a strong and non-evolving correlation between Ly$\alpha$ equivalent width (EW) and low-ionization interstellar absorption EW, revealing the connection between escaping Ly$\alpha$ photons \citep[and, by extension, ionizing radiation;][]{Reddy2016,Steidel2018,Chisholm2018a} and the covering fraction of neutral gas in the CGM. Building on these earlier works, we now seek to probe the rest-UV properties of galaxies at redshifts even closer to the epoch of reionization.

Approaching reionization at $z>6$, we must attempt to understand how the rest-UV spectrum reveals the changing properties of both massive stars and the ISM/CGM at higher redshift. The evolution of the relationship between Ly$\alpha$ emission strength and LIS absorption strength sheds light on the intrinsic production rate of Ly$\alpha$ photons in high-redshift galaxies, which in turn probes the intrinisic ionizing-photon production rate. 
Affecting observations near reionization, the IGM becomes more optically thick to Ly$\alpha$ photons and attenuates the spectra of galaxies. We must correct for this attenuation to accurately determine the changing intrinsic production rate of Ly$\alpha$ photons. As stated in D18, the Ly$\alpha$-LIS relationship also indicates the changing covering fraction of neutral gas in the CGM at higher redshift. Both intrinsic ionizing-photon production rate and neutral-gas covering fraction are key components of understanding the contribution of star-forming galaxies to the ionizing background during reionization \citep{Robertson2015a,Nakajima2018,Steidel2018,Trainor2019,Chisholm2018a,Chisholm2019}.

The DEIMOS 10K Spectroscopic Survey \citep{Hasinger2018} provides an ideal dataset for performing this type of analysis, including thousands of galaxies ranging from $z\sim0$ to $z\sim6$ with deep, rest-UV spectroscopy. This survey was conducted in the COSMOS field \citep{Scoville2006}, which is covered by extensive broadband photometry enabling the measurement of integrated galaxy properties.
In this work, we extend the analysis of D18 to higher redshift by carefully examining a $z\sim5$ sample of star-forming galaxies drawn from the DEIMOS 10K Spectroscopic Survey. We perform a joint analysis of the new $z\sim5$ sample with the $z\sim2\textrm{--}4$ samples of D18, using consistent methodology across all redshifts to remove any potential systematics.

In Section \ref{sample}, we introduce our sample of $z\sim5$ galaxies and the $z\sim2\textrm{--}4$ galaxies of D18, and discuss sample completeness, redshift measurements, and spectral energy distribution (SED) fitting. In Section \ref{methods}, we describe the methodology for measuring the equivalent widths of Ly$\alpha$ and interstellar absorption features, and the binning of the sample and subsequent creation of composite spectra. In Section \ref{results}, we present the resulting trends of rest-UV spectral features at $z\sim5$. In Section \ref{sec:disc_future} we analyze the sample at a finer redshift sampling and discuss the implications of our results for quantifying the changing intrinsic Ly$\alpha$ photon production and the contribution of star-forming galaxies to reionization. In Section \ref{sec:summary}, we summarize our key results and conclude.

Throughout this paper, we adopt a standard $\Lambda$CDM cosmology with $\Omega_m$ = 0.3, $\Omega_{\Lambda}$ = 0.7 and $H_0$ = 70 $\textrm{km\,s}^{-1}\textrm{Mpc}^{-1}$.

\section{Sample} \label{sample}

\subsection{Selection Criteria}

We analyzed a sample of galaxies drawn from the DEIMOS 10K Spectroscopic Survey of the COSMOS field \citep{Hasinger2018,Scoville2006}. This survey provides deep optical spectroscopy taken with the DEep Imaging Multi-Object Spectrograph (DEIMOS) on the Keck II telescope on Mauna Kea. The COSMOS field is covered by a wealth of ancillary data, including multi-wavelength imaging from space-based and ground-based missions.
Out of the $\sim$10,000 galaxies targeted in the DEIMOS 10K survey, we selected those identified as Lyman break galaxies (LBGs), dropping out in the $B_J$ and $g^+$ optical filters for $z\sim4$, and $V_J$ or $r^+$ for $z\sim5$. Also included were objects with photometric redshifts between $z\sim3.75\textrm{--}5.75$. These criteria yielded a parent sample of 416 star-forming galaxies at high-redshift, 337 selected via dropout and 79 via photometric redshifts.

We further required objects to have high-confidence spectroscopically-confirmed redshifts. 
The DEIMOS 10K catalog provided a quality flag $Qg$ indicating the confidence level of the spectroscopic-redshift measurement based on the quality of the spectrum and the number of features in the redshift estimate (either Ly$\alpha$ or LIS lines). 
We examined all objects with $Qg\geq1$ that satisfied the above photometric criteria, eliminating only those with unsuccessful redshift measurements. From this set of objects, we measured a redshift from the Ly$\alpha$ line, LIS lines, or both. We successfully recovered redshifts for 196 objects at $4<z<5.5$.  For the purpose of our completeness calculations, from our parent photometric sample of 416 galaxies, we removed galaxies selected by photometric redshift that had measured spectroscopic redshifts $z<4.0$, reducing the parent sample to 375.

\subsection{Systemic Redshifts} \label{sec:zsys}
\begin{figure}
        \includegraphics[width=\columnwidth]{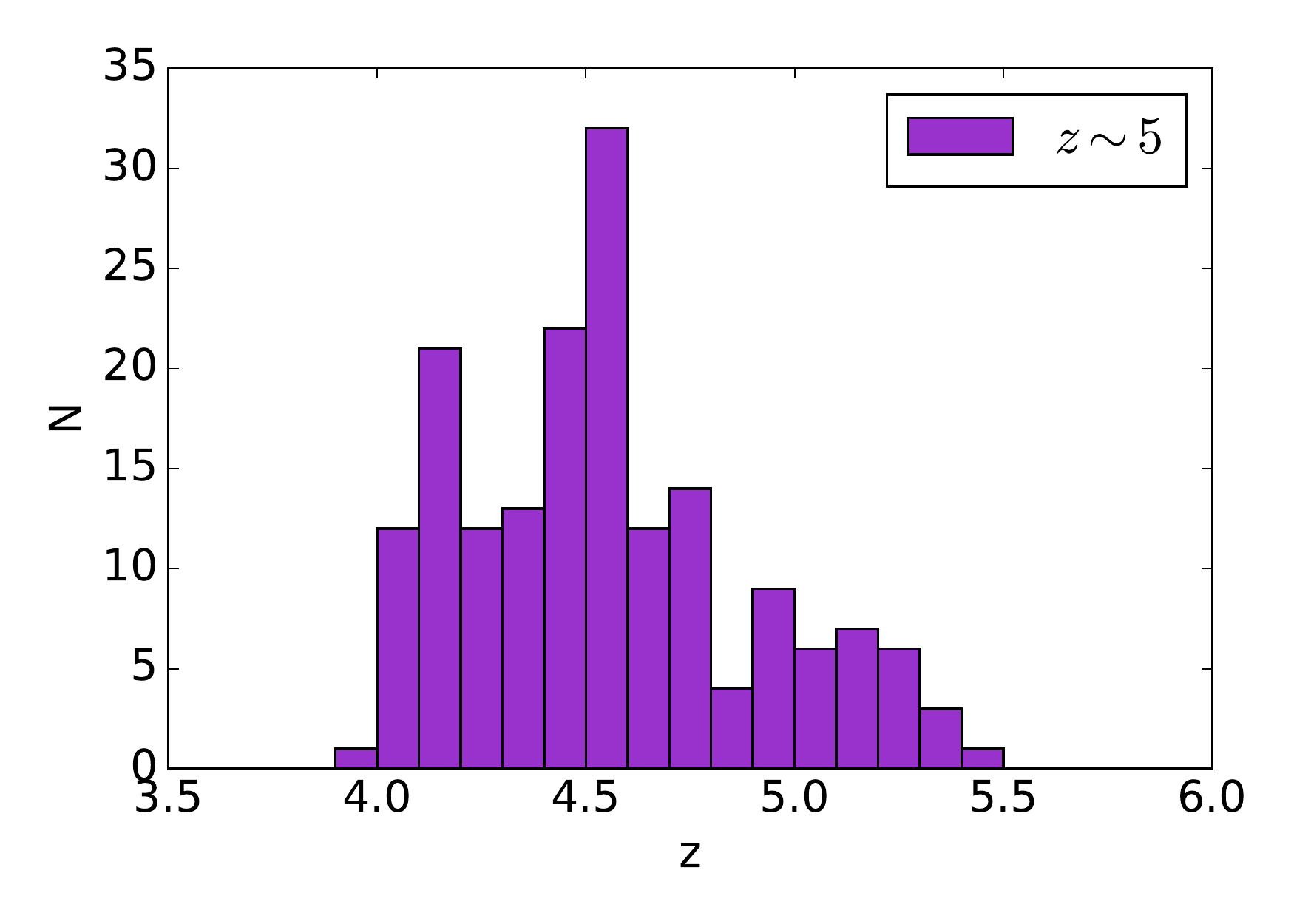}
	\caption{
		Redshift distribution of the DEIMOS 10K sample at $z\sim5$. The sample comprises 176 objects with a median redshift of 4.521.
	}
	\label{fig:zdist5}
\end{figure}

Ideally, rest-optical nebular emission lines would be used to estimate the systemic redshifts of high-redshift galaxies. As these lines are shifted into the thermal infrared at $z\sim5$, we measure the redshifts of the spectra via the redshifts of the Ly$\alpha$ emission line ($z_{\rm Ly\alpha}$) and the three strongest LIS absorption features ($z_{\rm LIS}$): Si~\textsc{ii}$\lambda1260$, O~\textsc{i}$\lambda1302+$Si~\textsc{ii}$\lambda1304$, and C~\textsc{ii}$\lambda1334$. Using the spectroscopic redshifts provided by DEIMOS 10K, we estimated the central wavelength of each non-resonant line and fit a Gaussian profile to the flux-calibrated spectrum. Each line was visually inspected to confirm the validity of the line detection and corresponding fit. The redshifts based on 
the central wavelengths of the well-fit Gaussian profiles of the three LIS lines were averaged with weights to calculate $z_{\rm LIS}$. Given the asymmetry of the Ly$\alpha$ profile, we did not model it with a Gaussian, but rather calculated $z_{\rm Ly\alpha}$ based on the wavelength at which the spectrum reached a maximum within the Ly$\alpha$ profile. We then determined the systemic redshift using the method described in \citet{rudie2012}. $z_{\rm Ly\alpha}$ was assumed to be offset from systemic by $300$~$\textrm{km\,s}^{-1}$ and $z_{\rm LIS}$ by $-160$~$\textrm{km\,s}^{-1}$, with these offsets empirically determined by \citet{Steidel2010}. While these rules are calculated at $z\sim2$, we assume they hold to $z\sim5$ in the absence of other information. This calculation enabled an estimate of the systemic redshift based on a high-quality measurement of either $z_{\rm Ly\alpha}$ or $z_{\rm LIS}$. If both $z_{\rm Ly\alpha}$ and $z_{\rm LIS}$ were measured, the systemic redshift was estimated as the average of the two. After combining 14 duplicate spectra, removing four spectra with poorly-fit lines and two spectra with background subtraction issues, we assembled a final sample of 176 objects between $4<z<5.5$: 68 with $z_{\rm Ly\alpha}$ only, 34 with $z_{\rm LIS}$ only, and 74 with both types of redshift measurement. The redshift distribution of the sample is displayed in Figure \ref{fig:zdist5}. The median redshift of our sample is $z_{\rm med}=4.521$, with a 16th--84th percentile range of 4.172--4.930. 
The redshift uncertainty $\sigma_z$ for each object was calculated using uncertainties on the assumed velocity offsets of Ly$\alpha$ and LIS, empirically determined in \citet{Steidel2010}. For galaxies with both $z_{\rm Ly\alpha}$ and $z_{\rm LIS}$ measured, we assumed $\sigma_{v_{\rm Ly\alpha}}$ = 175~$\textrm{km\,s}^{-1}$ and  $\sigma_{v_{\rm LIS}}$ = 115~$\textrm{km\,s}^{-1}$. For galaxies with only $z_{\rm LIS}$ measured, we assumed $\sigma_{v_{\rm LIS}}$ = 165~$\textrm{km\,s}^{-1}$. For galaxies with only $z_{\rm Ly\alpha}$ measured, we assumed $\sigma_{v_{\rm Ly\alpha}}$ = 175~$\textrm{km\,s}^{-1}$. 
The median $\sigma_z$ of the sample redshifts was 0.002.
We refer to this sample as the ``$z\sim5$ sample'' for simplicity and keep this sample distinct from samples featured in D18 to signify a new dataset.

\begin{figure}
	\includegraphics[width=\columnwidth]{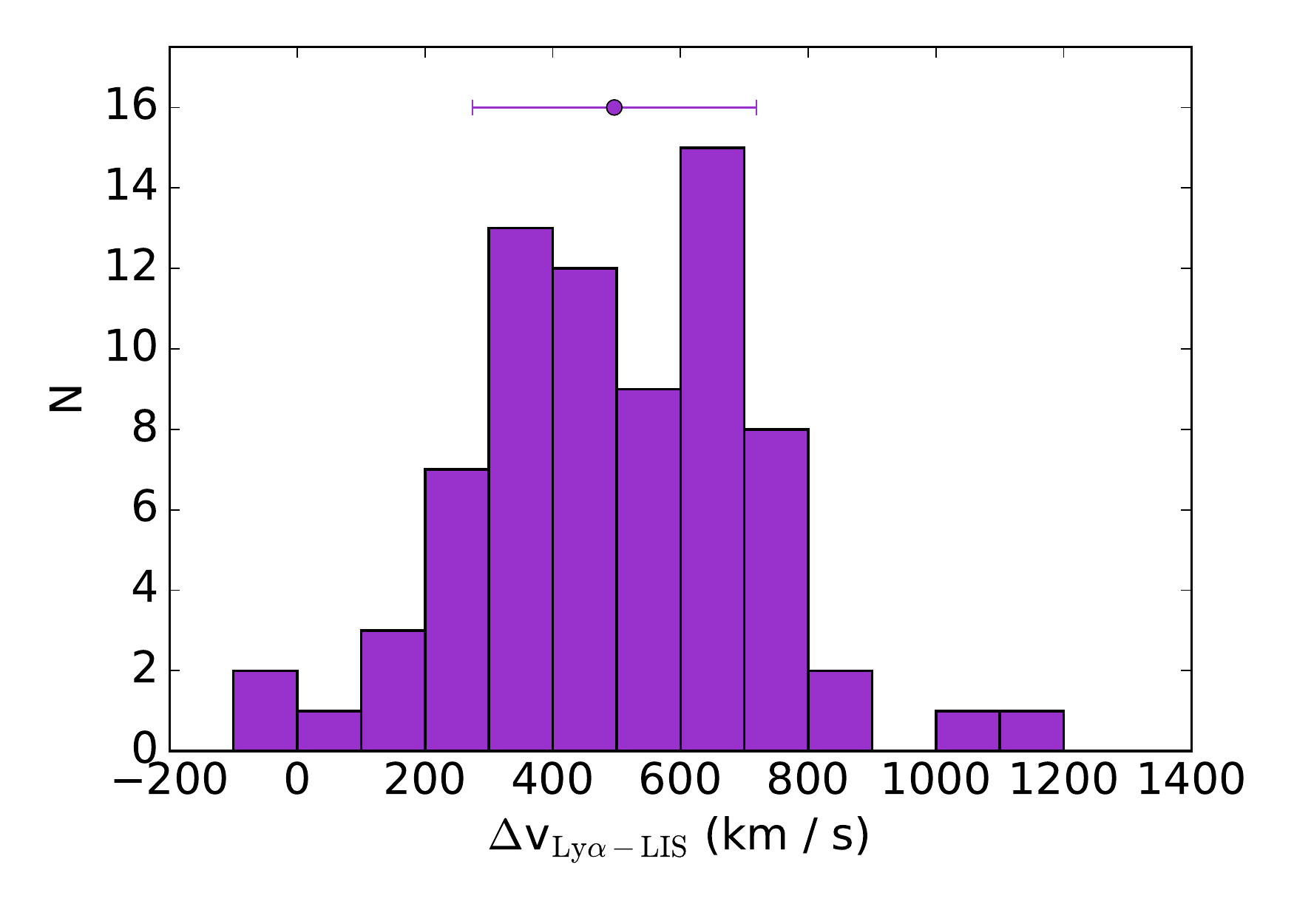}
	\caption{ Distribution of velocity offsets between Ly$\alpha$ emission and LIS absorption. The velocities are distributed about $\langle\Delta v_{\rm Ly\alpha - LIS}\rangle=496\pm222$~$\textrm{km\,s}^{-1}$.
	}
	\label{fig:veldist}
\end{figure}

For the objects with both Ly$\alpha$ and LIS absorption redshifts measured, we examined the distribution of velocity differences between Ly$\alpha$ and LIS lines in Figure \ref{fig:veldist}. 
This measurement was calculated as 
\begin{equation}
	\Delta v_{\rm Ly\alpha \textrm{--} LIS}= c \times \frac{(z_{\rm Ly\alpha} - z_{\rm LIS})}{(1 + (z_{\rm Ly\alpha} + z_{\rm LIS}) / 2)}.
\end{equation}
The velocity differences are distributed about a $\langle\Delta v_{\rm Ly\alpha \textrm{--}  LIS}\rangle=496$~$\textrm{km\,s}^{-1}$ with a standard deviation of 222~$\textrm{km\,s}^{-1}$. This value is largely consistent with $\langle\Delta v_{Ly\alpha  \textrm{--} LIS}\rangle\sim$ 600~$\textrm{km\,s}^{-1}$ of \citet{Steidel2010} at $z\sim2\textrm{--}3$, and $\langle\Delta v_{Ly\alpha \textrm{--} LIS}\rangle=429\pm229$~$\textrm{km\,s}^{-1}$ of \citet{Faisst2015} at $z\sim5$, the latter using a similar sample drawn from the DEIMOS 10K survey.
The distribution of velocity differences of Figure \ref{fig:veldist} has a similar shape to that of $z\sim3$ LBGs of \citet{Shapley2003}, a sample which has been shown to have similar outflow kinematics to those of the \citet{Steidel2010} $z\sim2$ sample. This similarity lends credence to our assumption that the outflow velocities calculated at $z\sim2$ are applicable out to $z\sim5$.

\subsection{Completeness}
\begin{figure}
	\includegraphics[width=.95\columnwidth]{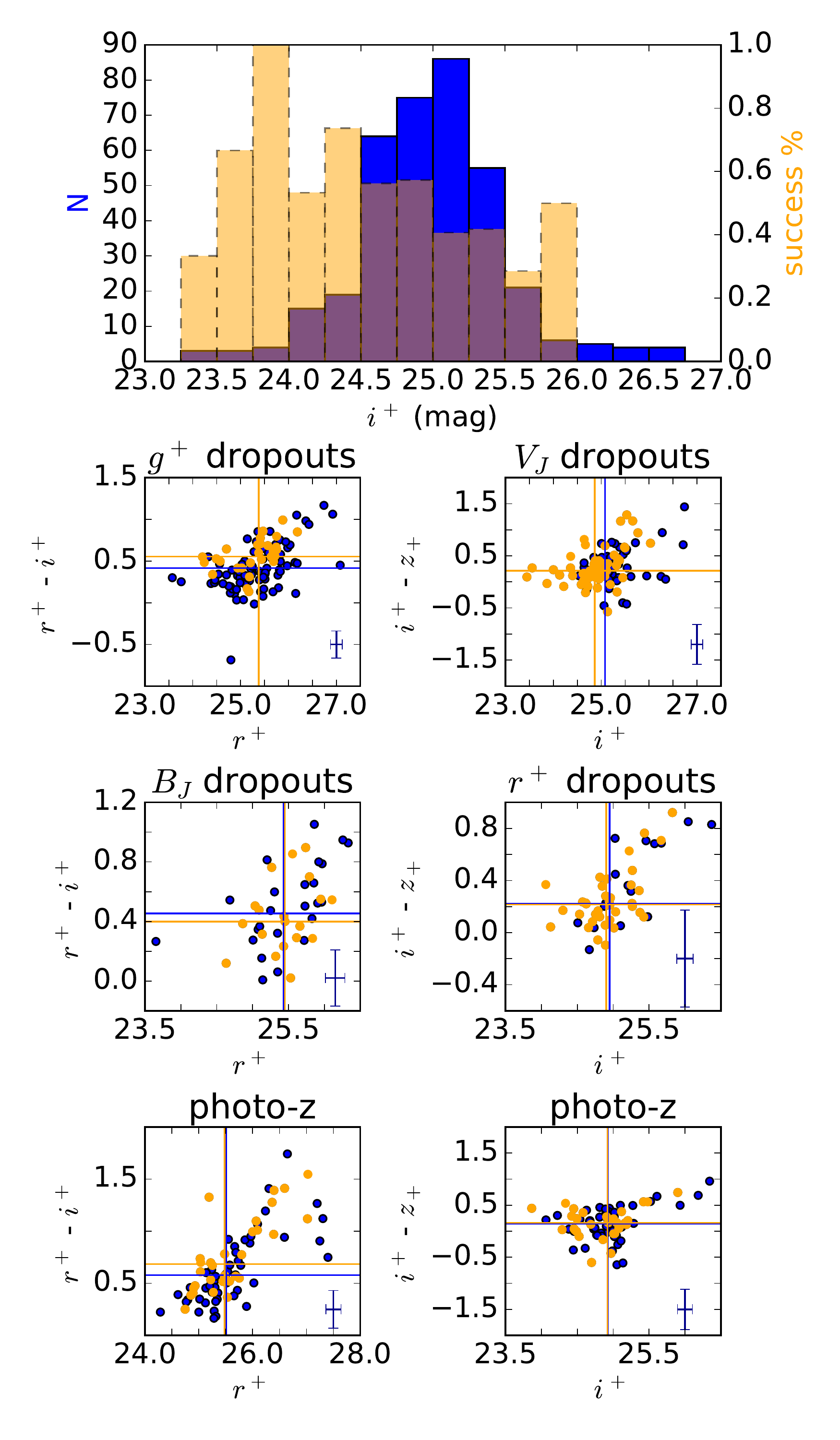}
        \caption{Completeness of the $z\sim5$ sample. 
        	\textbf{Top panel:} Histogram (in orange) showing the percentage of successfully measuring a high-confidence spectroscopic redshift in the $z\sim5$ parent sample as a function of $i^+$magnitude. The total $i^+$ magnitude distribution of the parent sample is overplotted in blue. The completeness drops below $\sim60$ per cent at an $i^+$ magnitude of $\sim$25.0.
        	\textbf{Middle four panels:} Color-magnitude diagrams of the dropout-selected galaxies. The parent sample is plotted in blue, with the corresponding median colors and magnitudes plotted as vertical and horizontal lines. The $z\sim5$ sample with well-measured redshifts is plotted in orange, along with the corresponding medians. Typical error bars for the respective samples are given in the lower-right corner of each panel. No significant deviation between the parent sample and the spectroscopic sample is found in color-magnitude space.
        	\textbf{Bottom two panels:} Color-magnitude diagrams of the galaxies selected by photometric redshifts between $3.75<z<5.75$. These panels follow the same color scheme as the dropout-selected galaxy panels.
        }
        \label{fig:completeness}
\end{figure}

In order to explore the completeness of the $z\sim5$ COSMOS DEIMOS sample, we examined the number of galaxies in the parent sample that report a high-confidence  redshift as a function of brightness. The results of this analysis are illustrated in the top panel of Figure \ref{fig:completeness}, where the total number of galaxies in the parent sample and the percentage of successful redshift measurements are plotted as a function of $i^+$-band magnitude. At $i^+ > 25$ mag, the sample completeness drops below $\sim60$ per cent, thus a more careful examination of the sample completeness must be performed.

As described above, spectroscopic redshifts at high redshift are measured using a combination of HI Ly$\alpha$ emission and LIS absorption lines of Si~\textsc{ii}$\lambda1260$, O~\textsc{i}$\lambda1302$+Si~\textsc{ii}$\lambda1304$, and C~\textsc{ii}$\lambda1334$. Measuring redshifts requires the robust detections of these lines. 
Since we explore the {\it variation} of the strengths of LIS and HIS absorption as a function of Ly$\alpha$ emission strength, strong detections should not bias our results. 
It is more important to examine whether the galaxies in our $z\sim5$ spectroscopic sample are representative of the parent photometric sample of star-forming galaxies in terms of other galaxy properties, such as color and brightness. To investigate this issue, we plotted our sample and the parent photometric sample in color-magnitude space in the bottom panels of Figure \ref{fig:completeness}. For the galaxies selected as dropouts, all filters chosen in the color-magnitude diagram are redward of the corresponding dropout filter. 
For both dropouts and photo-z selected samples, all but two of the deviations in median color are smaller than 0.06 mag, and all but one of the deviations in median magnitude are smaller than 0.05 mag. The lack of significant deviation in the median between our sample and the parent photometric sample demonstrates that the sample is not significantly biased in color and brightness.
The most significant deviation (0.22 mag) is between the median $i^+$-band magnitudes of the $V_J$ dropouts in our sample vs.~the parent sample, thus we examined our sample as a function of brightness in detail in section \ref{lyaew}. We conclude that our spectroscopic sample is not significantly biased towards certain types of star-forming galaxies with respect to the parent photometric sample in the same redshift range.

\subsection{$z\sim2\textrm{--}4$ sample}
\begin{figure}
	\includegraphics[width=\columnwidth]{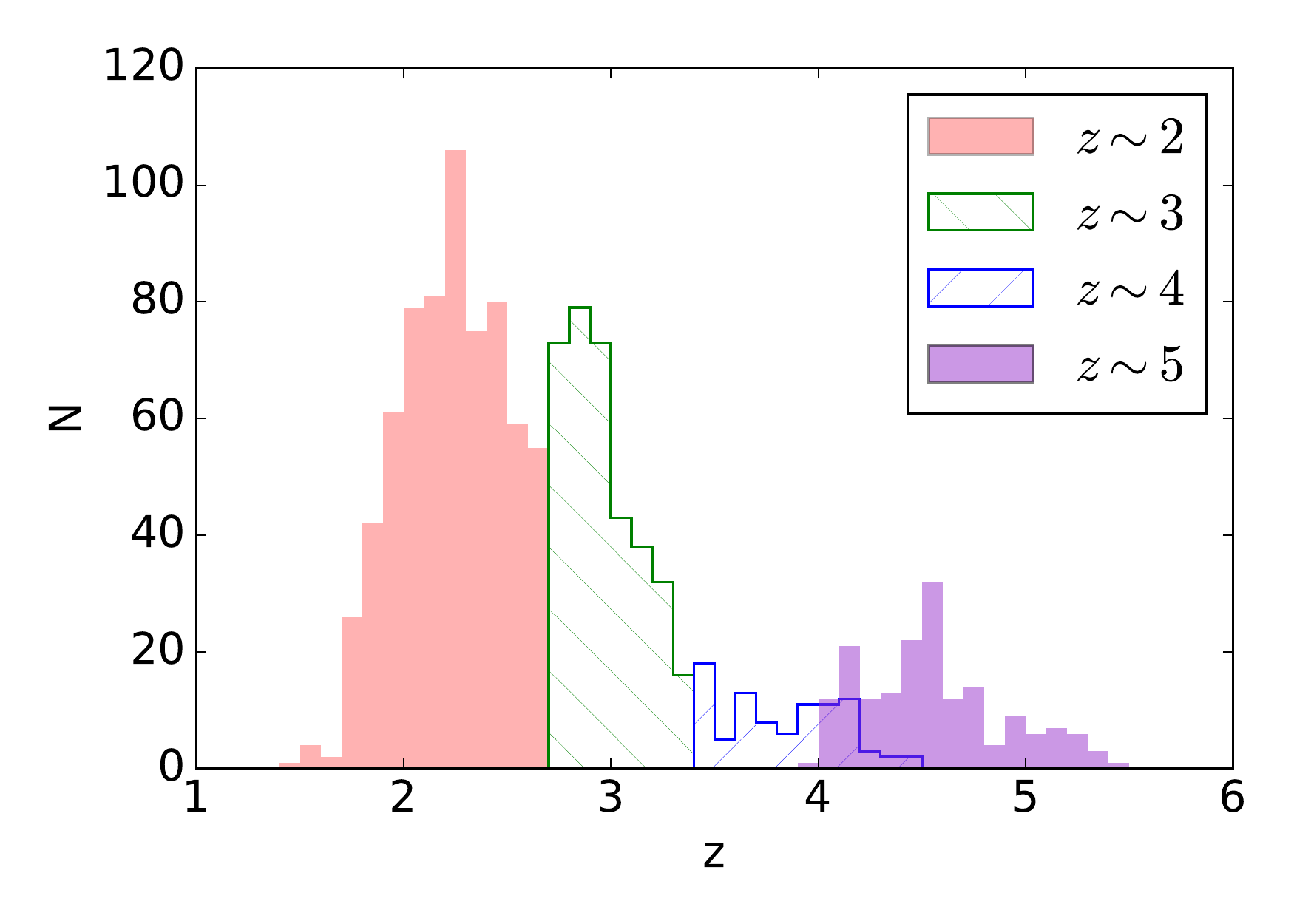}
	\caption{
		Redshift distribution of the four samples at $z\sim2\textrm{--}5$. The $z\sim2$, 3, and 4 samples in red, green, and blue are drawn from D18. The $z\sim5$ sample from this work is shown in purple.
	}
	\label{fig:zdist}
\end{figure}

For a lower-redshift comparison sample, we used the $z\sim2\textrm{--}4$ LBGs described in D18. The sample consists of $z\sim2\textrm{--}3$ galaxies preselected using $U_n GR$ color cuts and followed up with LRIS on Keck I as part of the redshift surveys described in \citet{Steidel2003}, \citet{Steidel2004}, and \citet{Reddy2008}. The $z\sim4$ sample was drawn from \citet{Jones2012} and consists of $B$-band dropouts in the GOODS-S field, 42 of which were followed up with DEIMOS on Keck II. In addition, 28 had spectra from the FOcal Reducer and low dispersion Spectograph 2 \citep[FORS2; ][]{Vanzella2004,Vanzella2006,Vanzella2007,Vanzella2009}  archival catalog. 
The average spectral resolutions were $R_{\rm average} \sim$ 970 and 1280 for the $z\sim2\textrm{--}3$ and $z\sim4$ samples respectively, compared to $R_{\rm average} \sim$ 2700 for the $z\sim5$ sample.
Systemic redshifts were measured in a similar method to that described above\footnote{There was a small difference in the method for estimating systemic redshifts in the case where both $z_{\rm Ly\alpha}$ and $z_{\rm LIS}$ were measured. D18 simply used $z_{\rm Ly\alpha}$ with an assumed velocity offset instead of averaging the two redshift measurements. In practice, the two estimates are very similar and this small difference in approach does not affect our results.}, and the sample was divided into three bins of increasing redshift, with 671 $z\sim2$ objects defined at $1.4\le z < 2.7$, 352 $z\sim3$ objects at $2.7\le z<3.4$, and 80 $z\sim4$ objects at $3.4\le z \le 4.5$. The full redshift distribution of the $z\sim2\textrm{--}5$ galaxies is displayed in Figure \ref{fig:zdist}. Some overlap does exist between galaxies in the $z\sim4$ and $z\sim5$ samples, but they are largely distinct with median redshifts of 3.856 and 4.521, respectively. In Section \ref{sec:disc_future}, we search for evolution within the $z\sim5$ sample by dividing this sample into two distinct redshift bins with $z_{\rm med}=4.330$ and 4.742. We note that the latter of these two more finely-sampled bins does not overlap in redshift space with the $z\sim4$ sample from \citet{Jones2012}.

\subsection{SED fits} \label{sed}
As we aim to compare spectral trends across a range of redshifts, we wanted to make a controlled comparison between similar types of galaxies at each redshift. To this end, we first made measurements of the galaxy stellar-population properties in each redshift bin via stellar-population synthesis model fits to the galaxy SEDs. We used photometry from the publicly-available catalogs of COSMOS2015 \citep{Laigle2016} for 171 of our objects, well matched in RA and Dec, and, for four objects, we used the photometric catalog of \citet{Capak2007}. For one object, we used the COSMOS Intermediate and Broad Band Photometry Catalog 2008\footnote{Available at https://irsa.ipac.caltech.edu/data/COSMOS/ datasets.html}, but, as this object did not have sufficient coverage redward of the Balmer break, we removed it from the sample, leaving us with 175 galaxies at $z\sim5$.

In our SED fitting, we utilized the Bruzual and Charlot (\citeyear{Bruzual2003}) stellar-population templates and assumed a Chabrier (\citeyear{Chabrier2003}) initial mass function (IMF). 
As these models do not include emission lines, we omitted the IRAC channel 1 band from the SEDs of galaxies at $4.00 < z < 4.99$ and the IRAC channel 2 band from the SEDs of galaxies at $5.06 < z < 5.31$. This correction removes potential SED contamination by the H$\alpha$ and [O~III]$\lambda5007$ rest-optical lines.
As in D18, we performed the SED fits using a combination of a fixed metallicity of 1.4 times the solar metallicity ($Z=0.02$) and a Calzetti extinction curve \citep{Calzetti1999}, or else a fixed metallicity of  0.28 $Z_{\odot}$ and an SMC extinction curve \citep{Gordon2003}. This method of SED fitting was initially described in \citet{Reddy2017}.
The star-formation history was set at either constant or exponentially increasing. The fits were performed with ages $>$ 10 Myr allowed, and then with only those $>$ 50 Myr. With these different assumptions, we fit for stellar mass (M$_{*}$), ages younger than the age of the universe, star-formation rate (SFR), and $E(B-V)$ from 0.00 to 0.60. Given that the quality of the fits of each model was similar, we adopted the best-fitting parameters of 0.28 $Z_{\odot}$, SMC attenuation, constant star-formation history, and age $>$ 50 Myr models. This choice was motivated by the fact that a sub-solar metallicity with an SMC extinction curve has been demonstrated to fit the IRX-$\beta$ relationship better than other models at $z>4$ \citep{Bouwens2016,Oesch2012}. 
Ages were constrained to $>$ 50 Myr so as to not be shorter than the typical dynamical timescales of distant star-forming galaxies \citep[assuming a relaxed disk and ignoring the timescsales of local gas motions,][]{Reddy2012}, and a constant SFR was assumed for consistency with past work \citep{Du2018,Reddy2012,Steidel2014,Strom2017}. 
The SED fitting of the $z\sim4$ sample in D18 was performed in a similar way. For the $z\sim2\textrm{--}3$ galaxies, the 1.4 $Z_{\odot}$+Calzetti model was used as they produced overall lower $\chi^2$.

\begin{figure*}
	\includegraphics[width=\textwidth]{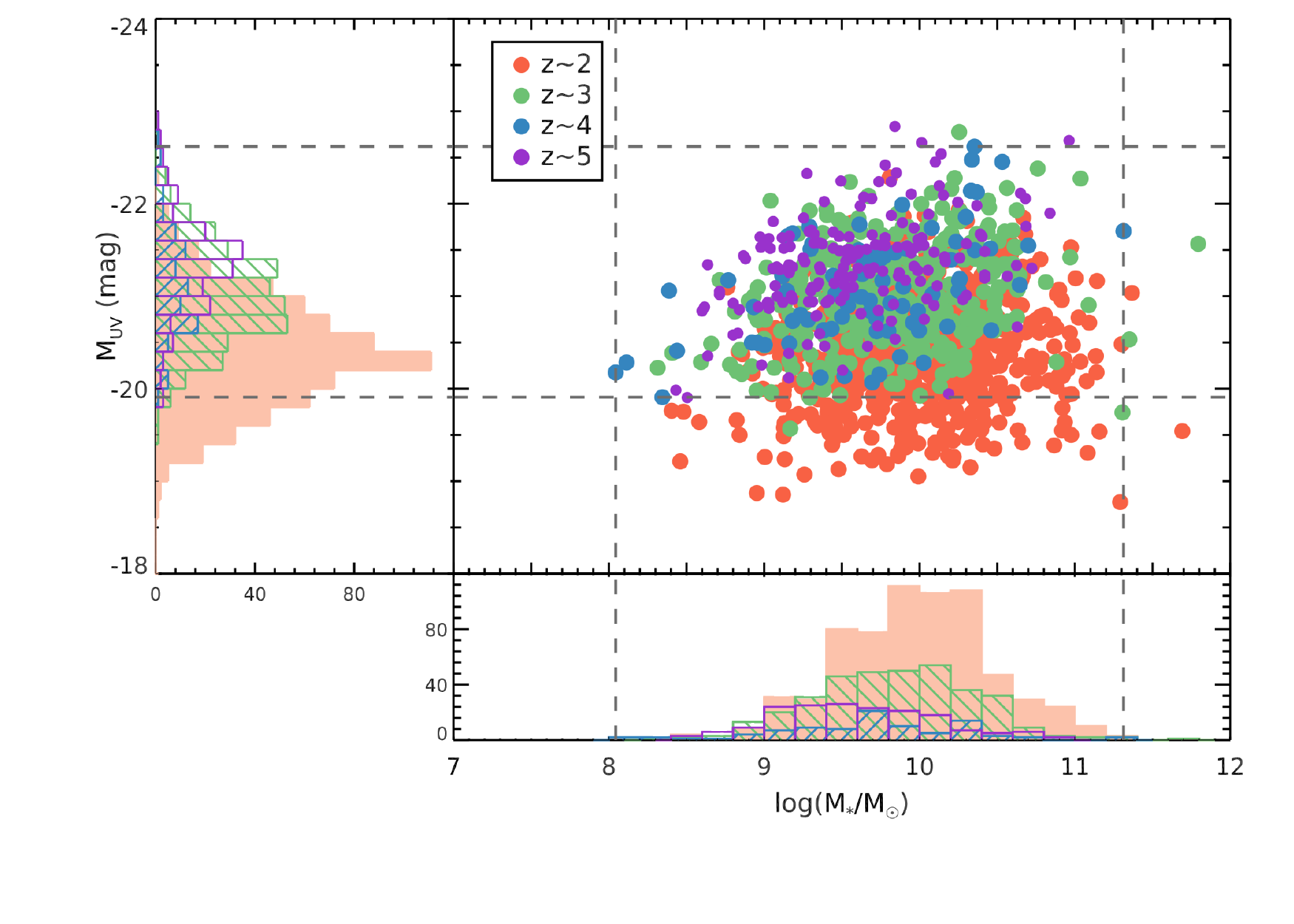}
	\caption{Absolute UV magnitude (M$_{\rm UV}$) vs. stellar mass (M$_{*}$) for the four redshift samples. The distributions of M$_{\rm UV}$ and M$_{*}$ are displayed in the histograms to the left and below the scatterplot. The cuts performed on the $z\sim2\textrm{--}4$ data are shown as vertical and horizontal dashed lines.
	}
	\label{fig:muv_m}
\end{figure*}

In order to perform a controlled study across redshift of galaxies with comparable properties, 
we restricted the $z\sim2$, 3, 4, and 5 samples to span the same range in stellar mass and rest-UV luminosity. We examined the distribution of M$_{*}$ determined by the SED fits and the UV absolute magnitude (M$_{\rm UV}$) in Figure \ref{fig:muv_m}. The M$_*-$M$_{\rm UV}$ cuts performed in D18 of  ($-22.62 < $ M$_{\rm UV} < -19.91$) and ($8.04 < $ log(M$_{*}$/M$_{\odot}) < 11.31$) are displayed as vertical and horizontal lines in the figure. These cuts remove low-luminosity objects in the $z\sim2$ sample.
The M$_{*}$ and M$_{\rm UV}$ of the $z\sim5$ sample span much of the same range as the $z\sim4$ sample. Since the $z\sim2\textrm{--}3$ samples were already truncated to match the $z\sim4$ sample, we perform no stricter M$_*-$M$_{\rm UV}$ cut on the $z\sim2$ and $z\sim3$ samples. The median M$_{\rm UV}$ and M$_*$ of each sample within the D18 cuts are recorded in Table \ref{tab:prop}, along with other median sample properties from the SED fitting. The similarity of the M$_{\rm UV}$ and M$_*$  medians further motivates that no additional cut is needed. After applying the D18 limits in M$_*$ and M$_{\rm UV}$, we find the number of objects in the $z\sim2$, 3, 4 and 5 samples are 539, 309, 91 and 175, respectively.

\begin{table}
\begin{center}
\begin{tabular}{llllll}
	\toprule
	{} & $z_{\rm med}$ & M$_{\rm UV}$ & log(M$_*$/M$_{\odot}$) & $E(B-V)$ & Sample Size \\
	\midrule
	$z\sim2$ &    2.267 &   -20.51 &                 10.00 &   0.09 &         539 \\
	$z\sim3$ &    2.925 &   -21.00 &                  9.87 &   0.08 &         309 \\
	$z\sim4$ &    3.856 &   -21.06 &                  9.72 &   0.04 &          91 \\
	$z\sim5$ &    4.521 &   -21.33 &                  9.80 &   0.05 &         175 \\
	\bottomrule
\end{tabular}
\end{center}
\caption{Median properties of the controlled $z\sim2\textrm{--}5$ samples.}
\label{tab:prop}
\end{table}

\section{Methods} \label{methods} 

Our goal is to examine the properties of the rest-UV spectral features of our different samples, including Ly$\alpha$; LIS lines of Si~\textsc{ii}$\lambda1260$, O~\textsc{i}$\lambda1302+$Si~\textsc{ii}$\lambda1304$, C~\textsc{ii}$\lambda1334$, and Si~\textsc{ii}$\lambda1527$; and HIS lines of Si~\textsc{iv}$\lambda\lambda1393,1402$ and C~\textsc{iv}$\lambda1548,1550$. While Ly$\alpha$ can be well detected at $z\sim5$ due to its large typical EW, it is difficult to detect the full suite of LIS and HIS lines in individual spectra, given their typical EWs of $\sim1-2\,$\AA{}. Thus, we rely on stacks of spectra in bins of various galaxy properties to reveal the characteristic strengths of the LIS and HIS lines. The Ly$\alpha$ line is known to have an important role in a galaxy's spectrum, strongly correlating with  rest-UV spectral morphology \citep{Shapley2003}, motivating its choice as a sorting parameter and the creation of composites in bins of Ly$\alpha$ EW. 
Ly$\alpha$ EW, as opposed to Ly$\alpha$ luminosity, is normalized to the UV continuum and probes the efficiency of Ly$\alpha$ photon production and escape.
The methods in this section broadly follow those described in D18. 
In addition, we consider here for the first time the increasing degree of IGM attenuation on average as a function of increasing redshift, quantifying this effect using the models of \citet{Laursen2010}.

\subsection{Ly$\alpha$ Equivalent Width in Individual Spectra} \label{lyaew}

In order to construct composite spectra spanning a range of Ly$\alpha$ EWs, we required an individual measurement of Ly$\alpha$ for each object. 
We used the fitting method of \citet{Kornei2010} to measure Ly$\alpha$ EW. First, the rest-frame spectra were examined individually to determine whether the spectrum at 1216\AA{} follows an ``emission," ``absorption," ``combination," or ``noise" profile. 
The large majority of our sample were fit using ``emission" or ``noise," representing either a positively-peaked, visible feature or no discernible one, with only one object following an ``absorption" profile consisting of a trough at 1216\AA{} and
four objects following a ``combination" profile of a blueside trough and redside peak.

The continuum level was then determined using the following procedure: 
the redside continuum was calculated as the average of the flux values of the spectrum between 1225\AA{}$\textrm{--}$1255\AA{}.
If the spectrum had coverage blueward of 1120\AA{}, the blueside continuum was estimated as the average of the flux between 1120\AA{}$\textrm{--}$1180\AA{}. If the spectrum only had coverage down to 1160\AA, the blueside continuum was averaged between 1160\AA{}$\textrm{--}$1180\AA{}. If the spectrum did not have coverage to 1160\AA{}, the blueside continuum value  was estimated by multiplying the redside value by the mean ratio of blue-to-red of 0.260, as determined from the remaining 140 objects of the sample with blueside coverage. 
The blueside continuum is suppressed relative to the redside due to absorption by the IGM.
For 14 of the objects, the redside continuum was not well detected, leading to a $\sim$zero continuum value well within the 1$\sigma$ noise of the spectrum. In order to measure an equivalent width for these galaxies, we estimated the continuum flux photometrically. This estimate was obtained using the best-fitting SED of the object as described in Section \ref{sed}, averaging in the same wavelength window of 1225\AA{}$\textrm{--}$1255\AA{}. These 14 spectra were also normalized over the wavelength interval 1300\AA{}$\textrm{--}$1500\AA{} so the continuum estimated from the spectra matched those from the photometry. For the objects with a well-detected continuum, this photometric estimation was also performed and compared to the spectroscopically-determined values to ensure consistency between the measurements.

\begin{figure}
 	\includegraphics[width=\columnwidth]{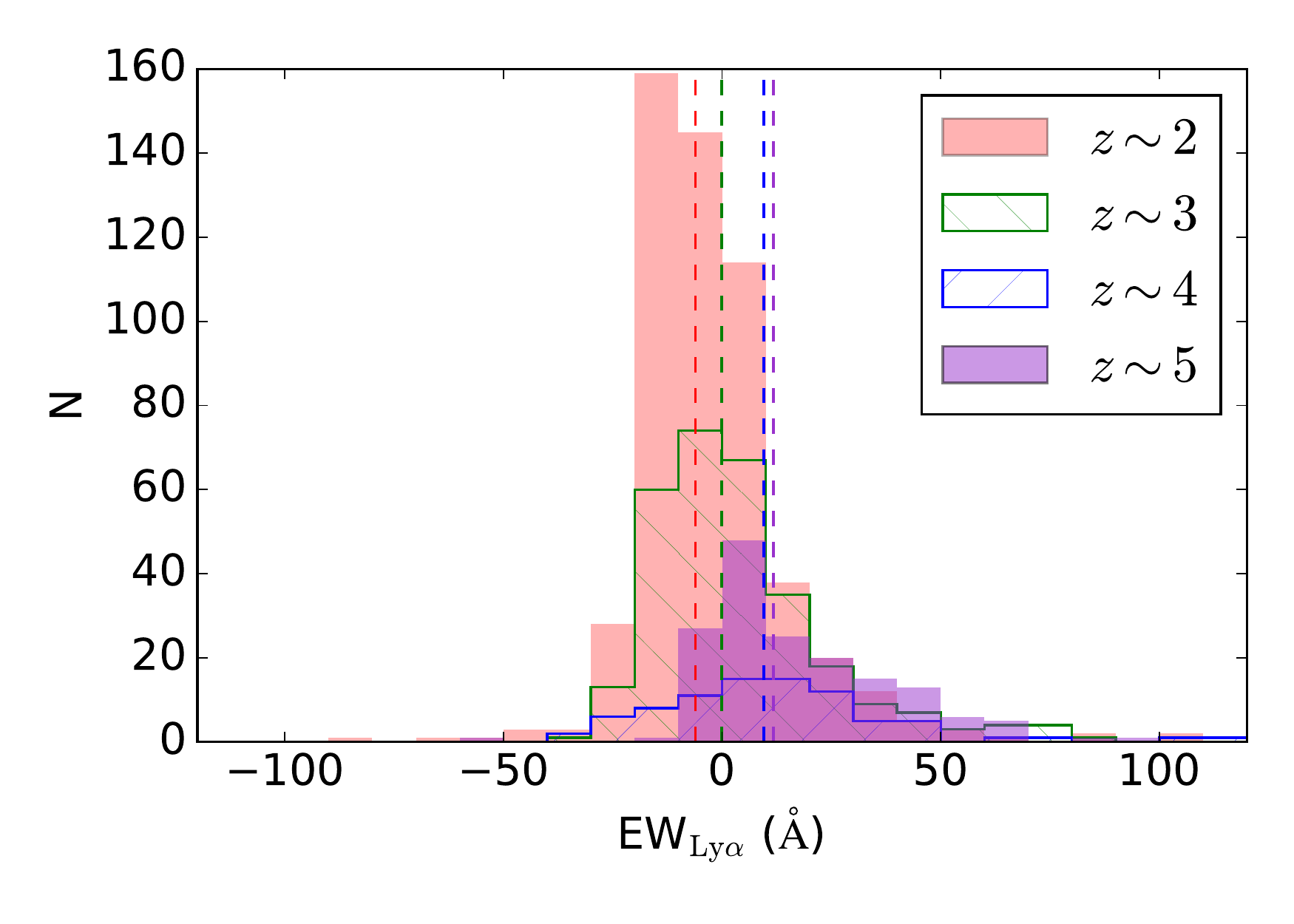}
        \caption{ Rest-frame EW$_{\rm Ly\alpha}$ distribution as a function of redshift. The $z\sim2$, 3, and 4 samples are from D18. The median EW$_{\rm Ly\alpha}$ for each sample is displayed as a vertical, dashed line. These medians increase towards stronger emission with increasing redshift.
        }
        \label{fig:ew_hist}
\end{figure}

For spectra classified as ``emission," ``combination," or ``absorption," the Ly$\alpha$ flux was integrated between the wavelengths at which the full Ly$\alpha$ profile intersected the blueside and redside continuum values. For ``noise," the spectrum was integrated over a fixed window of 1210$\textrm{--}$1220\,\AA{}. 
Finally, the equivalent width was measured as the ratio of the integrated Ly$\alpha$ flux and the redside-continuum flux density. The Ly$\alpha$ EW distribution of the four redshift samples is shown in Figure \ref{fig:ew_hist}. The median Ly$\alpha$ EW of each sample is overplotted on the corresponding histogram as a vertical line, indicating the evolution of the strength of Ly$\alpha$ within our four samples of increasing redshift. The sample median Ly$\alpha$ EWs are  $-6.07$\AA{}, $-0.08$\AA{}, 9.51\AA{}, and 11.75\AA{} for redshifts 2, 3, 4, and 5, respectively. This evolution must be interpreted in the context of the incompleteness of our $z\sim5$ sample.  
We find that 38.6 per cent of objects at $z\sim5$ had a systemic redshift measurement based only $z_{\rm Ly\alpha}$, compared to 56.3 per cent at $z\sim4$. In past work, a weak evolution of  Ly$\alpha$ emitting fraction has been found at these redshifts \citep{Stark2010}. 
We do not find a significant tail in the EW$_{\rm Ly\alpha}$ distribution at values less than 0~\AA\  (i.e., absorption) at $z\sim5$. However, we argue that, as the $z\sim5$ sample is not significantly biased in redshift measurement method or color-color space (Figure \ref{fig:completeness}), it is also not significantly biased with respect to the underlying Ly$\alpha$ EW distribution -- at least compared with $z\sim4$. See D18 for a discussion of sample completeness from $z\sim2$ to $z\sim4$.

\subsection{Composite Spectra} \label{sec:composite}

We first created composite spectra of the spectra in each redshift sample, and then split the samples into bins of various properties and generated composites of the spectra of the objects in each bin.

We generated composite spectra by performing a bootstrapping analysis of the objects in each sample. 
Based on the number of objects in the sample, we drew the same number of spectra from the bin with replacement. For each draw, we took each individual galaxy spectrum and normalized it to its photometric flux density. We then shifted it into the rest frame in $L_{\lambda}$ units based on its measured systemic redshift. The rest-frame, $L_{\lambda}$ spectrum of each object was then perturbed according to a Gaussian distribution with a standard deviation corresponding to the value of the error spectrum at that wavelength. These spectra were combined with the \textsc{iraf} tool $scombine$ using the combination mode ``average," with rejection of the three highest and three lowest flux values at each wavelength increment. This bootstrapping was performed 100 times, generating 100 composites, each representing a different sampling of spectra in the bin and reflecting the statistical noise in each individual spectrum. The resulting 100 spectra were then combined into a final science composite and error spectrum by taking the average $L_{\lambda}$ and standard deviation at each wavelength increment.

\begin{figure*}
	\includegraphics[width=\textwidth]{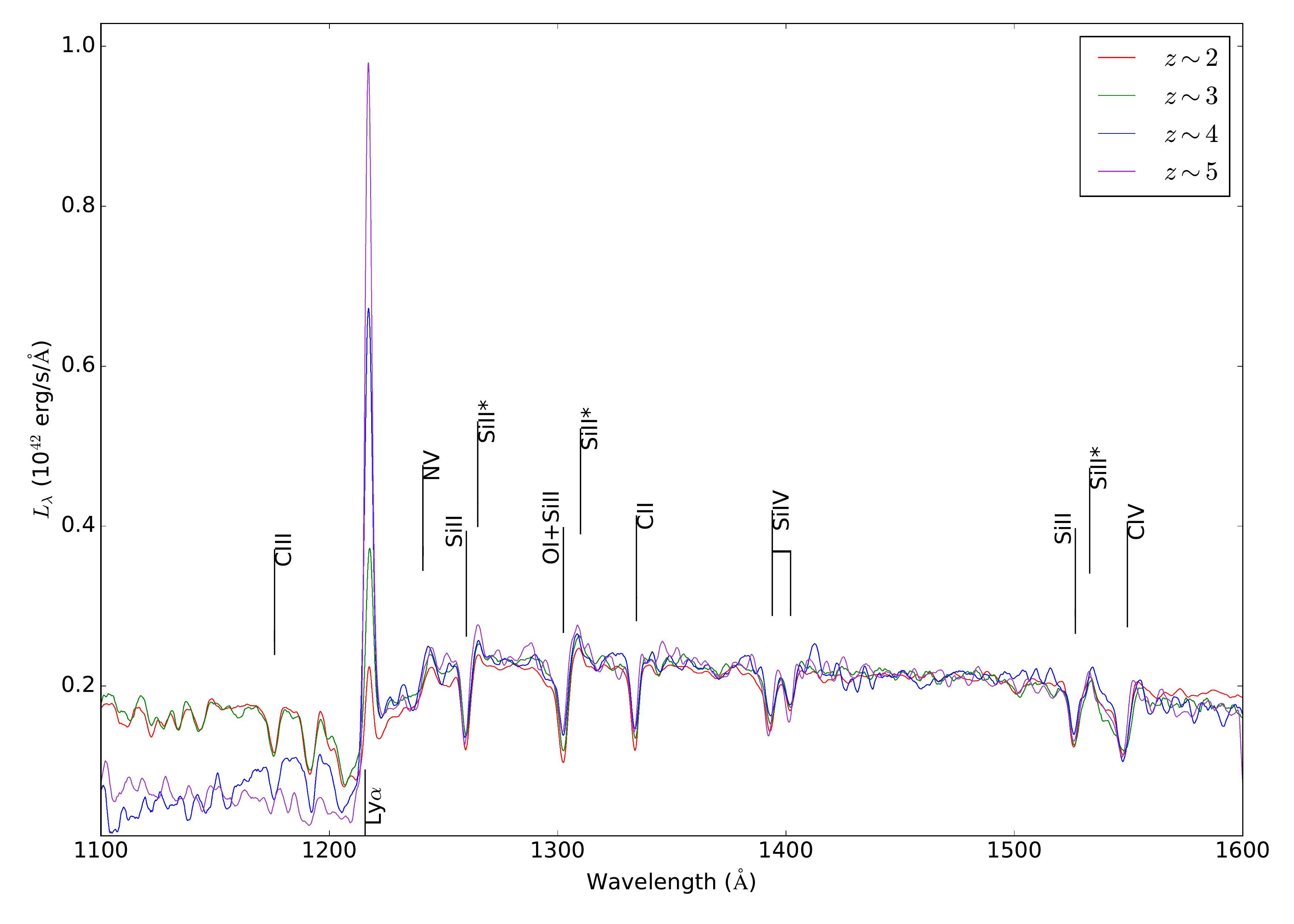}
	\caption{Composite spectra of the four redshift samples. These spectra were smoothed with a boxcar window of 3\AA{}. The $z\sim2\textrm{--}4$ spectra were normalized to the $z\sim5$ spectra using their respective median flux value in the range 1450-1500\AA{}. The rest-frame positions of various absorption and emission features are displayed.
	}
	\label{fig:full_composite}
\end{figure*}

We can examine the overall spectral characteristics of each sample with this composite method. The four composites of increasing redshift are presented in Figure \ref{fig:full_composite}. The most striking difference in the $z\sim5$ spectrum is the strength of its Ly$\alpha$ feature, reflecting the difference in median Ly$\alpha$ EW of the sample. The properties of the absorption features are difficult to distinguish qualitatively and require further analysis.

We split our redshift sample into bins of various properties. In addition to Ly$\alpha$, these properties include $E(B-V)$ and integrated galaxy properties such as M$_*$, SFR, age, and M$_{\rm UV}$. Depending on the property being examined, the set of objects comprising each bin change based on the spectral coverage of the objects. For examining LIS strength as a function of Ly$\alpha$ EW, we required every object in the sample to have coverage out to the most redward LIS line, Si~\textsc{ii}$\lambda1527$. This requirement slightly reduced our sample to 160 objects. We split this sample into four bins of increasing Ly$\alpha$ EW with 40 objects per bin. Using an equal number of objects per bin provides a comparable S/N for each composite spectrum. To examine HIS strength as a function of Ly$\alpha$ EW, we required coverage of the most redward HIS line, C~\textsc{iv}$\lambda1548,1550$. We divided the resulting sample of 154 objects into four bins of increasing Ly$\alpha$. Finally, we examined Ly$\alpha$ and LIS strength as a function of integrated galaxy properties using four bins of the 160 objects with coverage of Si~\textsc{ii}$\lambda1527$.

\begin{figure*}
 	\includegraphics[width=\textwidth]{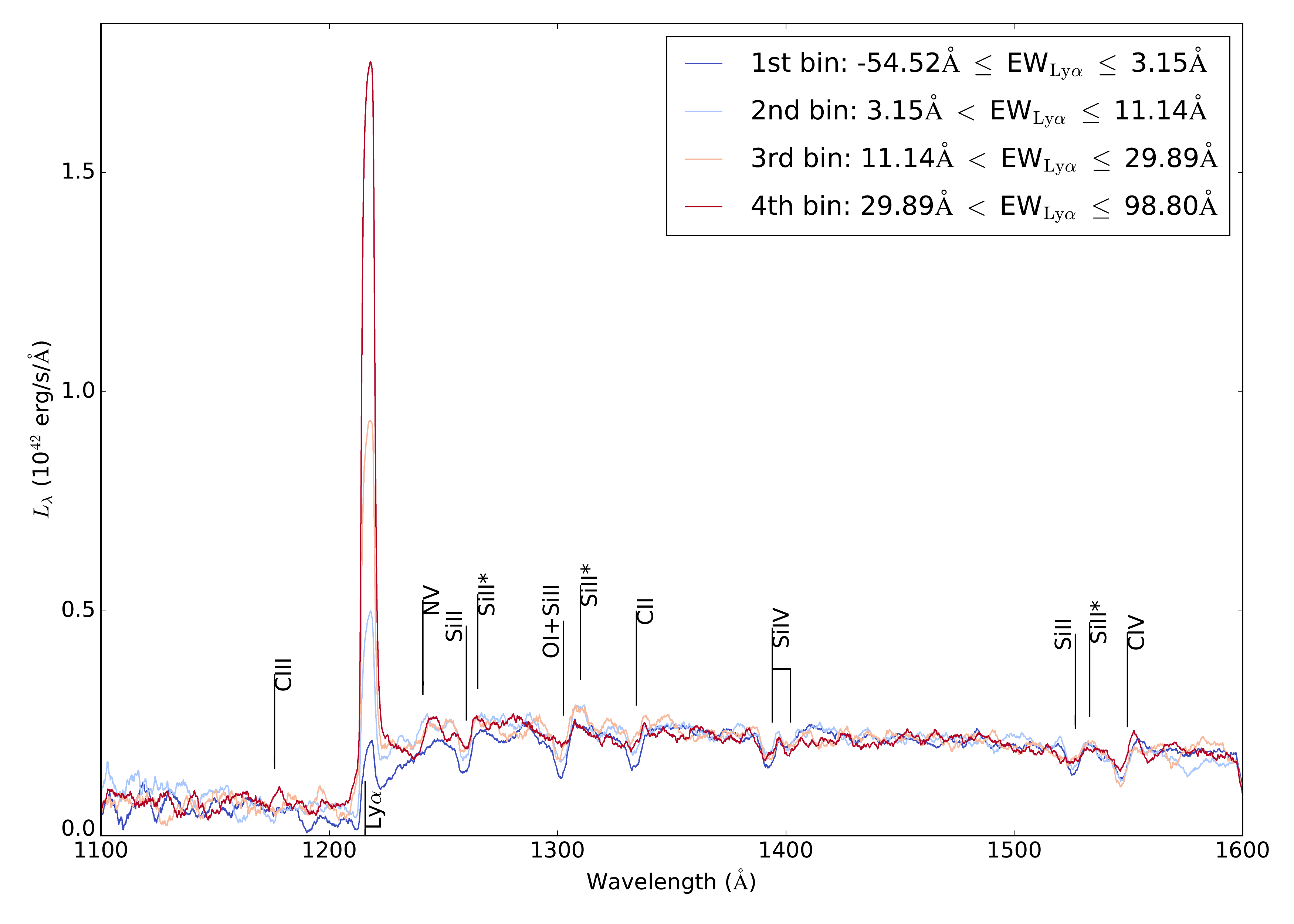}
        \caption{ Composite spectra of the $z\sim5$ sample in bins of increasing Ly$\alpha$ EW. The dark blue spectrum corresponds to the objects with the lowest EW$_{\rm Ly\alpha}$, while the dark red contains the objects with the highest EW$_{\rm Ly\alpha}$. Each composite spectrum is constructed from the spectra of 34 objects characterized by the EW$_{\rm Ly\alpha}$ ranges displayed in the legend. The variation of rest-UV spectral features as a function of increasing EW$_{\rm Ly\alpha}$ can be qualitatively observed near the wavelengths of various LIS and HIS lines labelled by solid black lines.
        }
        \label{fig:composites_bins}
\end{figure*}

The four composite spectra binned by Ly$\alpha$ EW are shown in Figure \ref{fig:composites_bins}. The absorption features are clearly resolved in the composite spectra, and the variation of the Ly$\alpha$ EW in each bin can be seen in the profile at 1216\AA{}. From these composites, we can remeasure the Ly$\alpha$ EW and measure the EW of interstellar absorption features. To determine the Ly$\alpha$ EW of each bin, we applied the fitting technique described in Section \ref{lyaew} to each of the bootstrapped spectra, representing the variation in each composite bin of Figure \ref{fig:composites_bins}. The final Ly$\alpha$ EW and 1$\sigma$ error reported for each bin was the mean and standard deviation of these 100 measurements. This measured Ly$\alpha$ EW from stacked spectra and the median Ly$\alpha$ EW of the objects composing each bin were comparable within 1$\sigma$.

\begin{figure}
  \includegraphics[width=\columnwidth]{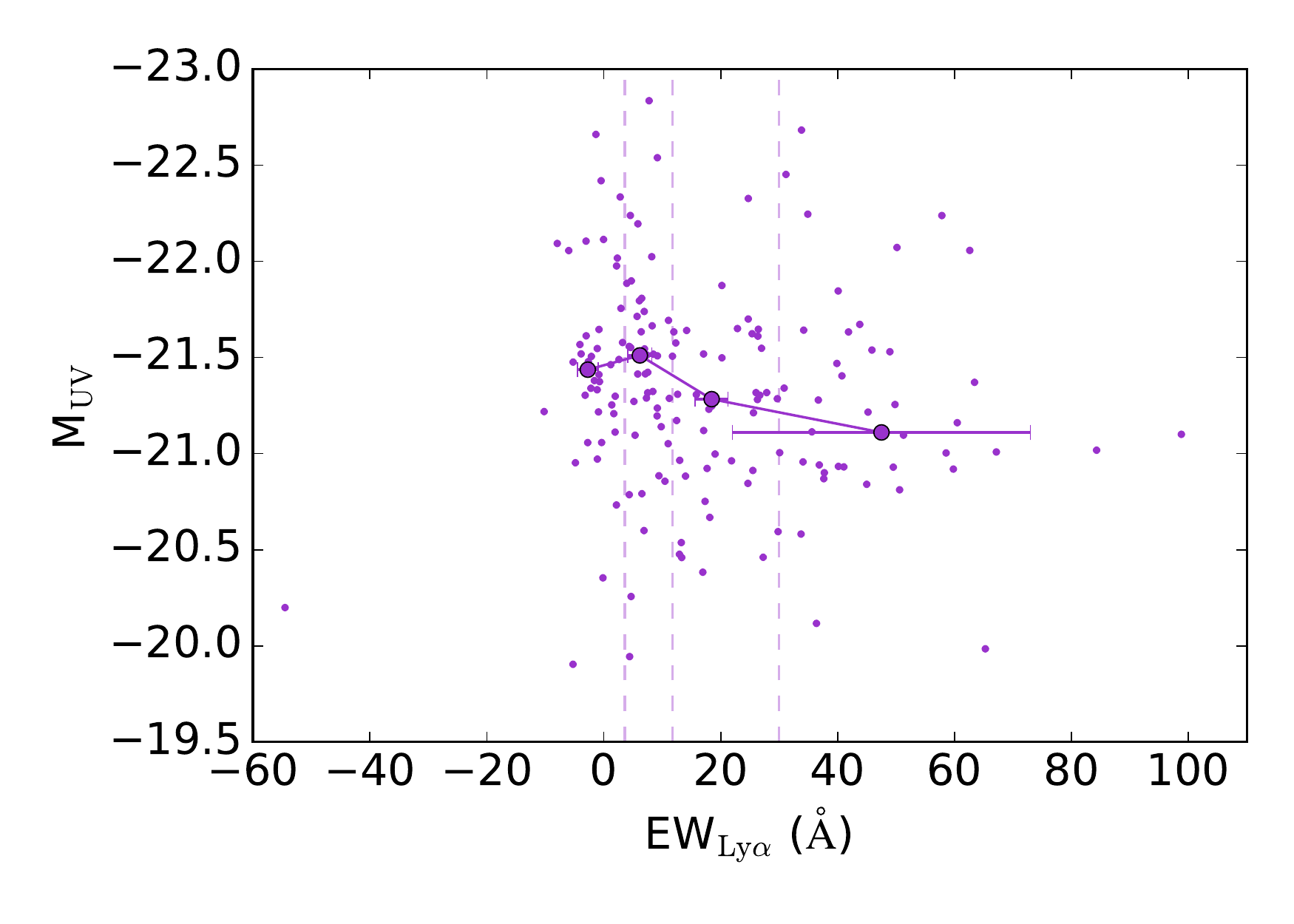}
  \caption{ Absolute UV magnitude (M$_{\rm UV}$) vs. EW$_{\rm Ly\alpha}$. Individual detections are displayed as smaller circles. The sample is divided into bins of increasing Ly$\alpha$ EW, with these divisions shown as vertical, dashed lines. The median properties of the objects of each bin are plotted as larger, filled circles.
}
  \label{fig:muv_lya}
\end{figure}

To understand how our analysis is affected by the changing intrinsic brightness of the galaxies, we examined the M$_{\rm UV}$ of each galaxy as a function of its measured Ly$\alpha$ EW. All the galaxies in our $z\sim5$ sample are plotted in M$_{\rm UV}$-EW$_{\rm Ly\alpha}$ space in Figure \ref{fig:muv_lya}. As Ly$\alpha$ emission strength increases, the galaxies tend to be dimmer in the UV continuum. The objects are split into four bins of increasing Ly$\alpha$ EW, and the median M$_{\rm UV}$ and Ly$\alpha$ EW are displayed. The median M$_{\rm UV}$ of the lowest EW$_{\rm Ly\alpha}$ bin is $-21.44$, while the highest bin has a median M$_{\rm UV}$ of $-21.11$. While we observe a trend of fainter M$_{\rm UV}$ with increasing EW$_{\rm Ly\alpha}$, just as that found within the $z\sim3-4$ samples of D18, this dependence is weak. Accordingly, it is reasonable to interpret differences in EW$_{\rm Ly\alpha}$ as being dominated by variations in emergent Ly$\alpha$ flux.

\subsection{IGM Correction to Ly$\alpha$ Equivalent Widths} \label{sec:igm}

In order to obtain a more accurate measurement of the intrinsic Ly$\alpha$ production and escape through the ISM, we attempted to remove the effects of IGM attenuation on the strength of the observed Ly$\alpha$ line.
The IGM transmission in the vicinity of the Lya feature decreases noticeably from $z\sim2$ to $z\sim5$. To quantify this evolution, \citet{Laursen2010} generated sightlines through high-resolution cosmological simulations of galaxy formation, where the simulated sightlines originate in samples of galaxies at different redshifts ranging from $z=2.5$ to $6.5$. These authors produced the mean IGM transmission profile in the vicinity of Ly$\alpha$ at each redshift, showing how its shape and depth evolve towards stronger absorption at higher redshift. We interpolated the model of \citet{Laursen2010} to produce IGM transmission curves appropriate for the median redshift of each of our four redshift samples. \footnote{We note that each IGM transmission curve has an associated uncertainty based on the range of sightline opacities found in the simulations of \citep{Laursen2010}. Given that this uncertainty for the average of $\sim$40 sightlines is subdominate compared to our other uncertainties, we did not take it into account.}

\begin{figure}
	\centering
	\includegraphics[width=0.5\textwidth]{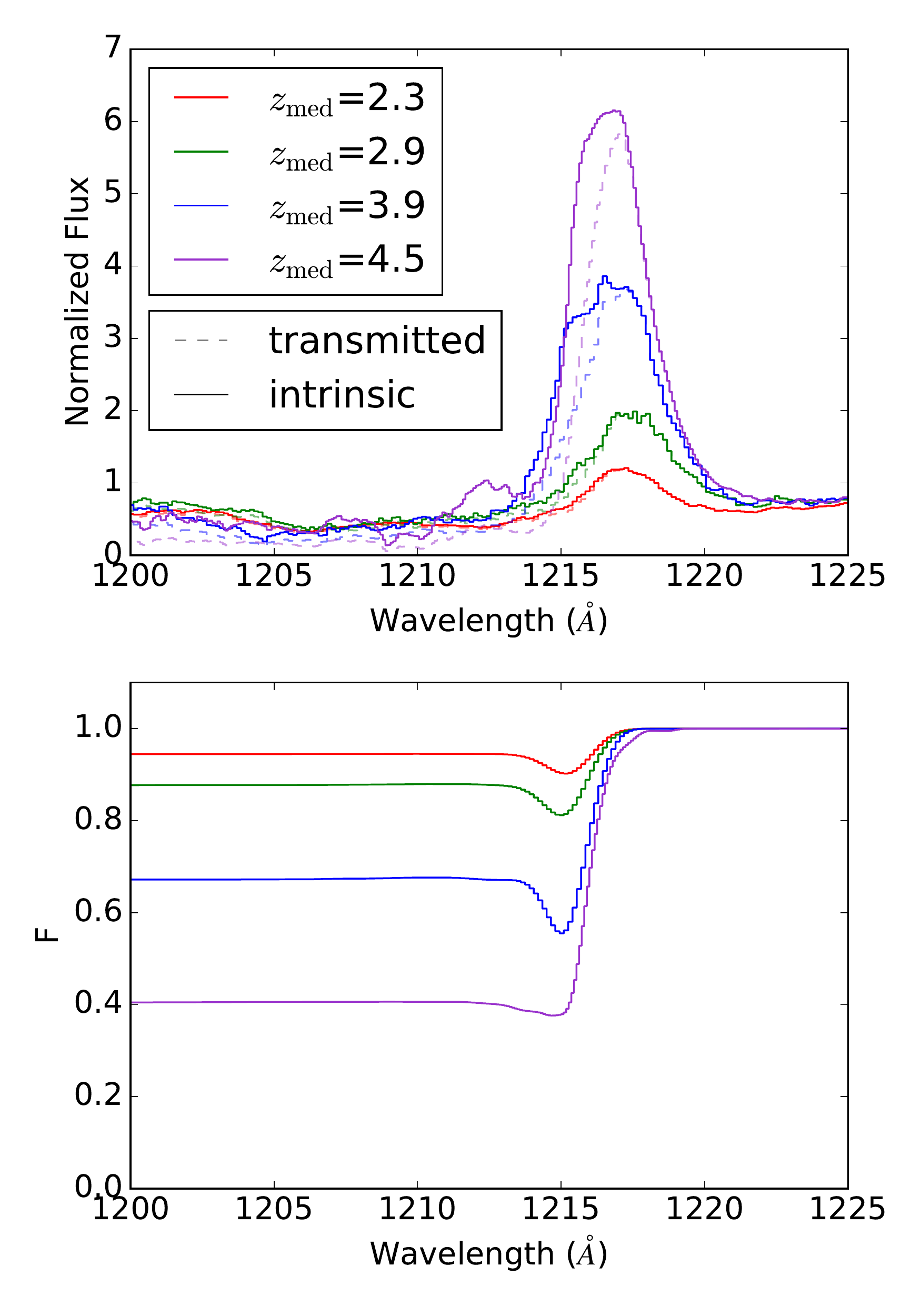}
	\caption{ IGM correction method applied to the $z\sim2\textrm{--}5$ composites spectra.
		\textbf{Top panel:} Ly$\alpha$ profiles for $z\sim2\textrm{--}5$ sample composites, with and without IGM attenuation correction applied. The observed composites transmitted through the IGM are displayed as dashed lines, and the intrinsic Ly$\alpha$ profiles are displayed as solid lines. The intrinsic profiles were determined by dividing the transmitted profiles by the transmission curves shown in the bottom panel.
		\textbf{Bottom panel:} Degraded IGM transmission curves calculated at $z=2.3, 2.9, 3.9$, and 4.5. Transmission curves were calculated at a given redshift by interpolating the models of \citet{Laursen2010} and were degraded based on the average spectral resolution of each redshift sample and the systemic redshifts of the objects contained in each sample.
	}
	\label{fig:igm}
\end{figure}

We then degraded the curves in a manner that represented the process of observing our galaxy spectra at moderate resolution and combining them with uncertain systemic redshifts into composite spectra. Specifically, we smoothed the model IGM transmission curve by a Gaussian kernel to match the average spectral resolution of each composite: $R_{\rm average} \sim$ 970, 1280, and 2700 for the $z\sim2\textrm{--}3$, $z\sim4$, and $z\sim5$ samples respectively. The more important blurring effect arises from the fact that the estimates of systemic redshift from Ly$\alpha$ and LIS features have associated systematic errors (i.e., they are based on mean relations between Ly$\alpha$ or LIS redshifts and systemic redshifts, which have intrinsic scatter). In order to simulate the effect of stacking N spectra with uncertain systemic redshifts, we generated N individual IGM transmission curves and perturbed each curve by a small $\Delta \lambda$. The value $\Delta \lambda$ was drawn from a normal distribution with zero mean and a standard deviation of $\sigma_{\Delta \lambda}=(\sigma_z / (1 + z)) \times 1215.67$\AA{}. 
The redshift uncertainty $\sigma_z$ was calculated through the method described in Section \ref{sec:zsys}.
The resulting N perturbed (in wavelength) transmission curves were then averaged together. The transmission curves calculated in this way for the $z \sim 2 - 5$ samples are shown in the bottom panel of Figure \ref{fig:igm}. The fraction of photons transmitted by the IGM at 1215.67\AA{} decreases substantially towards $z \sim 5$.

In order to convert the transmitted Ly$\alpha$ profiles to those intrinsic to the galaxy, we divided the observed composite spectra by the associated degraded IGM transmission curve. Both the transmitted and intrinsic Ly$\alpha$ profiles of the $z \sim 2 - 5$ samples are shown in the top panel of Figure \ref{fig:igm}. The Ly$\alpha$ profiles in composite spectra at higher redshift receive larger positive corrections for IGM absorption. This analysis was applied to the composites of the $z \sim 2 - 5$ samples binned by Ly$\alpha$ EW and galaxy properties, and the Ly$\alpha$ EW of each composite was remeasured as described in section \ref{sec:composite}.

\subsection{Absorption Line Equivalent Width} \label{sec:absew}

\begin{figure*}
 	\includegraphics[width=\textwidth]{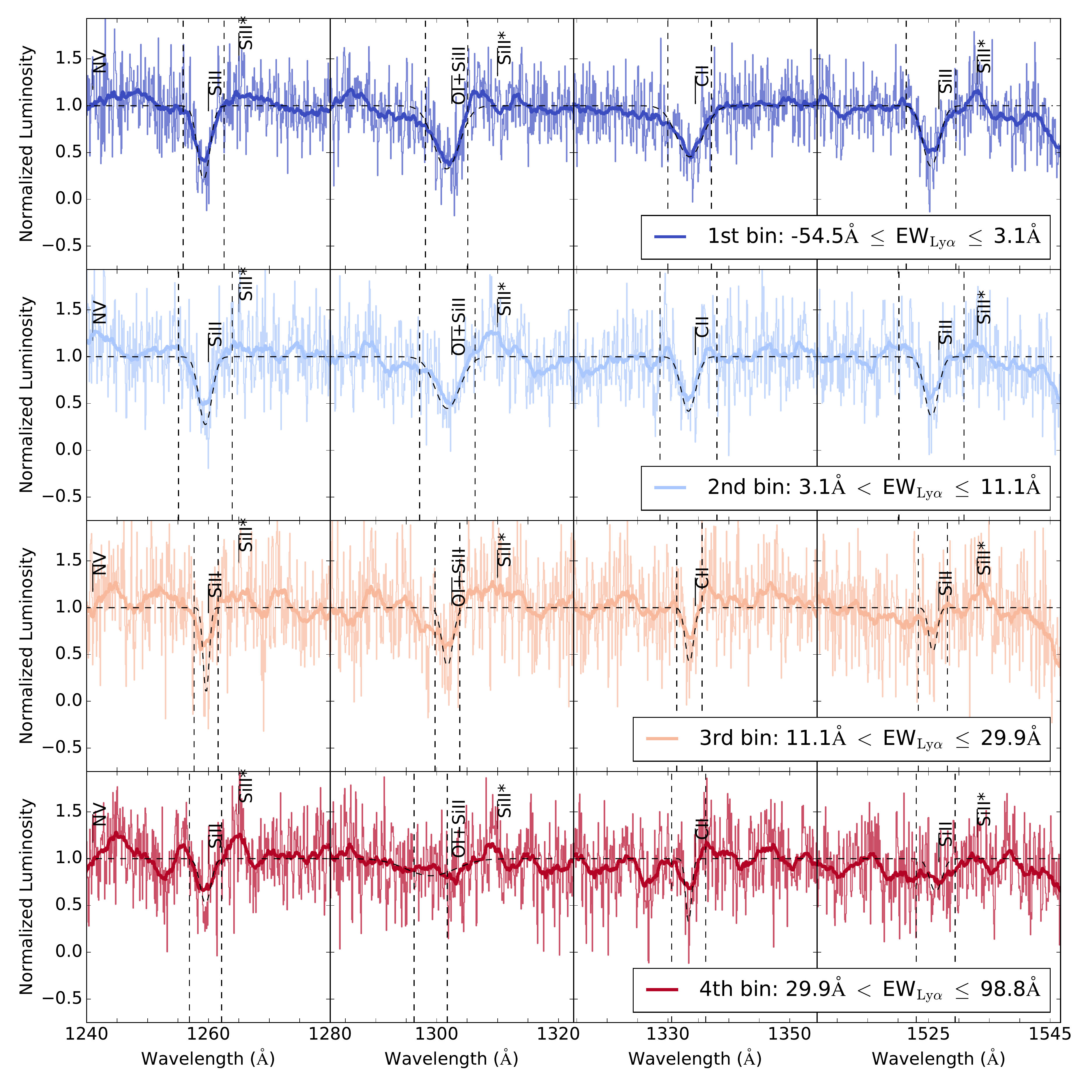}
        \caption{Fits to the four LIS features of the composite spectra of the $z\sim5$ sample, split into four bins with EW$_{\rm Ly\alpha}$ decreasing from the top to the bottom row. These EW$_{\rm Ly\alpha}$ bins are the same as those shown in Figure \ref{fig:composites_bins}. The more transparent spectra in each row are the continuum-normalized, rest-frame, $L_{\lambda}$ spectra from which measurements were performed. The more opaque spectra are convolved with a boxcar window of width 3\AA{} for easier viewing. The vertical, dashed lines  illustrate the $\pm5$\AA{} integration windows from which EW measurements were made.
        }
        \label{fig:lis_fit}
\end{figure*}

For each composite spectrum, we measured both LIS and HIS equivalent widths. 
First, we normalized the continua of the rest-frame, $L_{\lambda}$ spectra using the IRAF tool $continuum$, using a fourth-order spline fit with continuum regions that avoid the spectral features defined by \citet{Rix2004}.	
From the normalized spectra, we then fit each absorption feature with a Gaussian profile using the scipy routine $leastsq$, with initial fit parameters corresponding to the rest-frame central wavelength of the line and a linewidth of $\sigma=2$\AA{}, typical of composite LIS lines at $z\sim5$. 
To calculate the flux of each line, we integrated the composite spectrum about the central wavelength of the Gaussian fit to the line using an integration window that spanned from $-3\sigma_v$ to $+3\sigma_v$. Here, $\sigma_v$ is defined as the standard deviation in velocity space of the average of the four key LIS profiles under consideration. In order to determine $\sigma_v$, we converted the four LIS profiles from rest-frame wavelength space to velocity space and averaged them at each velocity increment. We then fit a Gaussian function to the combined absorption profile in velocity space using the scipy routine $leastsq$, leading to an estimate of $\sigma_v$. This technique yielded an integration window for each LIS feature that was informed by the average width of the LIS lines in the composite spectrum. At the same time, averaging the four LIS features in velocity space mitigated the effects of low-quality individual profile fits.
We finally calculated the EW of the absorption feature by dividing this flux by the normalized continuum value in the integration window. The error on the EW measurement was determined via propagation of errors from the error spectrum in the integration window.
The continuum-normalized spectra in bins of increasing Ly$\alpha$ EW are shown in Figure \ref{fig:lis_fit}. The fits to the Gaussian profiles of the LIS lines of Si~\textsc{ii}$\lambda1260$, O~\textsc{i}$\lambda1302+$Si~\textsc{ii}$\lambda1304$, C~\textsc{ii}$\lambda1334$, and Si~\textsc{ii}$\lambda1527$ are displayed in detail here. The integration windows of $\pm$5\AA{} are also displayed, with the entirety of the line profiles fitting in these windows.

\begin{figure}
  \centering
	\includegraphics[width=0.5\textwidth]{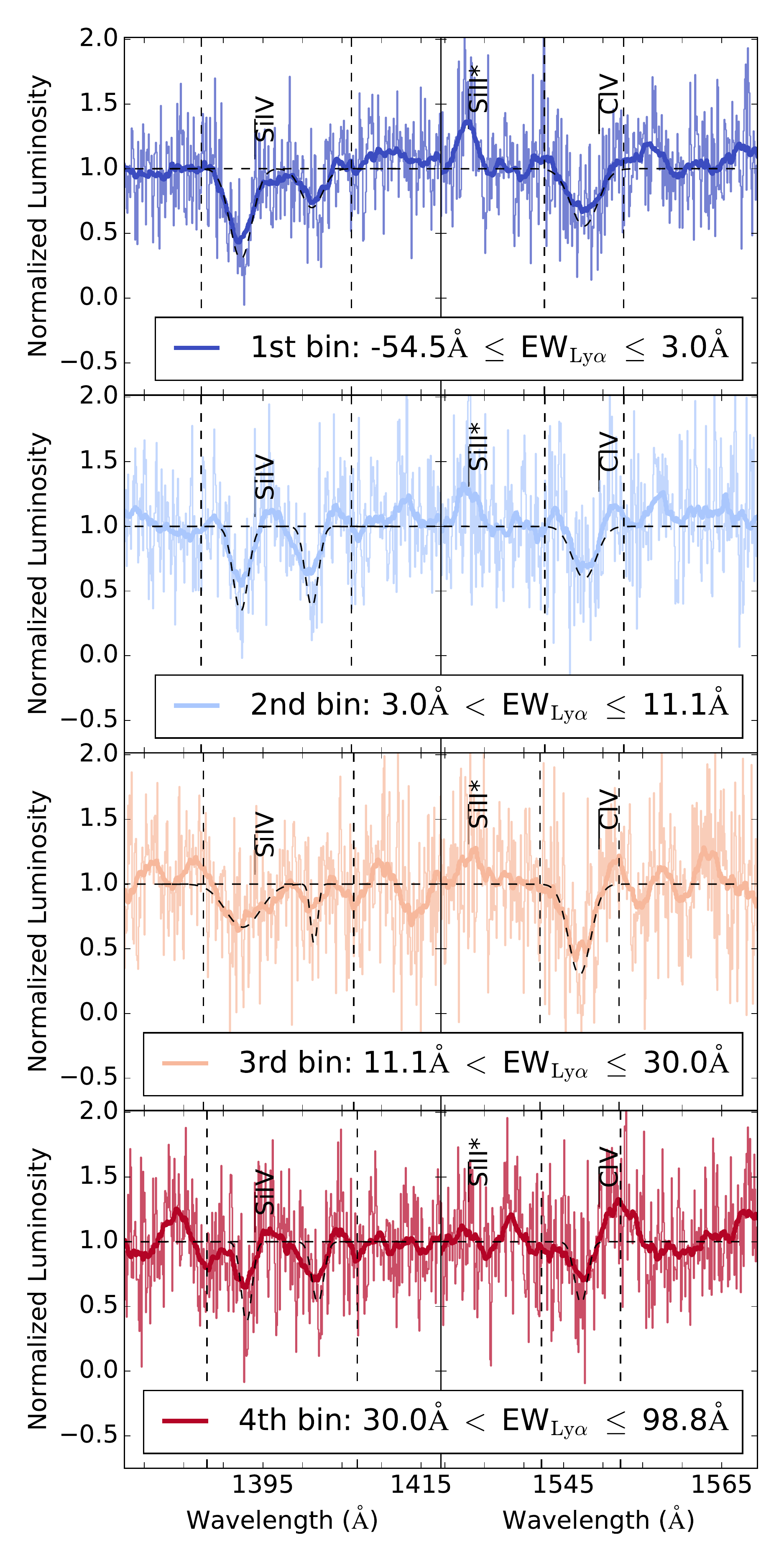}
        \caption{Fits to the Si$\:$IV$\lambda\lambda1393,1402$ and C~\textsc{iv}$\lambda1548,1550$ features of the composite spectra of the $z\sim5$ sample, split into four bins of increasing EW$_{\rm Ly\alpha}$. These spectra have the stellar component of  C~\textsc{iv}$\lambda1548,1550$ removed. The top panel corresponds to the objects with the lowest EW$_{\rm Ly\alpha}$, while the bottom panel contains the objects with the highest EW$_{\rm Ly\alpha}$. As in Figure \ref{fig:lis_fit}, the transparent spectra are the continuum-normalized, rest-frame, $L_{\lambda}$ spectra from which measurements were performed. The opaque spectra are convolved with a boxcar window of width 3\AA{} for easier viewing. The vertical, dashed lines  illustrate the $\pm5$\AA{} integration windows from which EW measurements were made.
        }
        \label{fig:his_fit}
\end{figure}

For the HIS doublet of Si~\textsc{iv}$\lambda\lambda1393,1402$, a double Gaussian profile was fit to the line in the same method described for the single profiles, above. Additionally, the central-wavelength ratio of the doublet members was fixed at the theoretical value of (1402.770/1393.755). The integration window was set to range from $-5$\AA{} of the centre of the bluer profile up to +5\AA{} of the centre of the redder profile. These Si~\textsc{iv}$\lambda\lambda1393,1402$ fits in four bins of increasing EW$_{\rm Ly\alpha}$ are shown in the left panels of Figure \ref{fig:his_fit}.

The C~\textsc{iv}$\lambda1548,1550$ profile is more complex: on top of the blended absorption feature, there is a stellar component produced by high-velocity winds in O and B stars \citep[][and references therein]{Du2016}. This stellar component must be removed before the absorption from the interstellar gas can be measured.
While a stellar P-Cygni component can appear on the Si~\textsc{iv}$\lambda\lambda1393,1402$  absorption feature, the component is less significant within the mass ranges and corresponding metallicity ranges of our samples \citep{Steidel2014,Sanders2018}, thus does not require modeling and removal. 
We used the rest-frame UV model starburst spectra of \citet{Leitherer2010}, smoothed to the resolution of our data, to represent the stellar component.
We fit the templates, evaluated at 0.05 $Z_{\odot}$, 0.2 $Z_{\odot}$, 0.4 $Z_{\odot}$, $Z_{\odot}$, and 2 $Z_{\odot}$, and combined the two bracketing (or one adjacent for the lowest- and highest-metallicities) best-fitting models. 
This fit is performed on the blue wing of C~\textsc{iv}$\lambda1548,1550$ between 1535 and 1544\AA{}, and we refer readers to Figure 3 of \citet{Du2016} for an illustration of our fitting method.
By dividing the continuum-normalized spectra by the best-fitting stellar template, we effectively removed the stellar component of C~\textsc{iv}$\lambda1548,1550$. We then measured the EW of the interstellar component of C~\textsc{iv}$\lambda1548,1550$ using the method previously described for the single LIS lines, as the doublet is blended together into a single profile at our spectral resolution. The fits to the C~\textsc{iv}$\lambda1548,1550$ interstellar profile after removing the stellar component are shown in the right panels of Figure \ref{fig:his_fit}.

\section{Results} \label{results} 

In this section, we examine the the properties of our $z\sim5$ sample as a function of binned properties of Ly$\alpha$ EW, $E(B-V)$, stellar mass, M$_{\rm UV}$, age, and SFR. 

\subsection{Line Strength}

We are motivated by a physical picture in which the strength of LIS absorption probes the covering fraction of neutral gas in a galaxy \citep{Reddy2016}. Thus,
with the reasonable assuption of a roughly spherically symmetric gas distribution \citep[e.g., ][]{Law2012},
there will be a connection between the Ly$\alpha$ photons, which are scattered by any neutral gas covering the galaxy \citep{Steidel2010}, and the resulting strength of the LIS absorption features. In contrast, there is evidence that the HIS lines exist in an ionized medium, physically distinct from the neutral gas \citep{Shapley2003,Du2016,Du2018}. 
HIS features have been found to have similar kinematics as LIS lines \citep{Pettini2002,Heckman2015,Chisholm2016,Berg2018}, however \citet{Du2016} finds that while the kinematics of HIS and LIS lines are similar, they correlate differently with galaxy properties, indicating they may exist in different ISM phases. 
A lack of relationship between Ly$\alpha$ and HIS will support this picture.

\begin{figure*}
  \centering
 	\includegraphics[width=\textwidth]{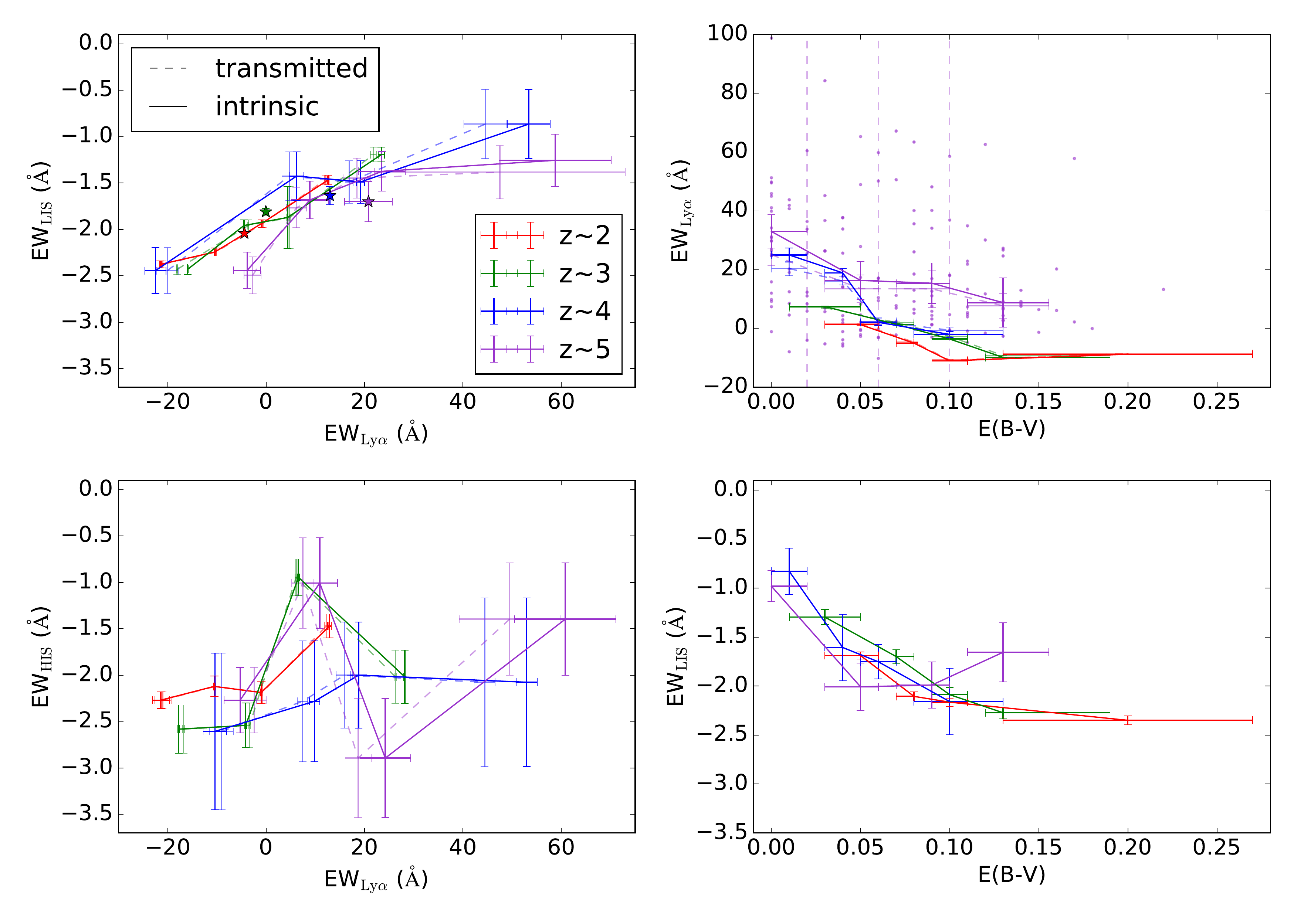}
        \caption{
        	Relations between EW$_{\rm Ly\alpha}$, EW$_{\rm LIS}$, EW$_{\rm HIS}$, and $E(B-V)$. EW$_{\rm Ly\alpha}$, EW$_{\rm LIS}$, and EW$_{\rm HIS}$ were measured from composite spectra. The measurements from the observed composites are displayed in transparent points connected by dotted lines, while the corrected measurements after application of degraded IGM transmission curves are the opaque points connected by solid lines. The purple points correspond to measurements of the $z\sim5$ sample. The red, green, and blue points are of the $z\sim2$, 3, and 4 samples of D18. 
        	\textbf{Top left panel:} EW$_{\rm LIS}$ vs. EW$_{\rm Ly\alpha}$. The redshift samples are divided into four bins of increasing EW$_{\rm Ly\alpha}$, with EW$_{\rm Ly\alpha}$ and EW$_{\rm LIS}$ measured from composite spectra composed of the objects in each bin. The average measurements of the $z\sim2\textrm{--}5$ samples are displayed as stars.
        	\textbf{Bottom left panel:} EW$_{\rm HIS}$ vs. EW$_{\rm Ly\alpha}$. The binned samples are the same as in the top left panel, with the exception that the objects were required to have coverage of C~\textsc{iv}$\lambda1548,1550$. EW$_{\rm HIS}$ and EW$_{\rm Ly\alpha}$ are also measured from composite spectra, made from combining the individual spectra of the objects in each EW$_{\rm Ly\alpha}$ bin.
        	\textbf{Top right panel:} EW$_{\rm Ly\alpha}$ vs. $E(B-V)$. The points are the redshift samples divided into four bins of increasing $E(B-V)$, with EW$_{\rm Ly\alpha}$ measured from composite spectra and $E(B-V)$ from the median of the objects in each bin. The transparent points are individual detections of the $z\sim5$ sample.
        	\textbf{Bottom right panel:} EW$_{\rm LIS}$ vs. $E(B-V)$. The binned samples are the same as the top right panel, with EW$_{\rm LIS}$ also measured from composite spectra.
        }
        \label{fig:lis_his_ebmv_lya}
\end{figure*}

To quantify the strength of the LIS lines, we measure EW$_{\rm LIS}$ as the weighted average of the Si~\textsc{ii}$\lambda1260$, O~\textsc{i}$\lambda1302+$Si~\textsc{ii}$\lambda1304$, C~\textsc{ii}$\lambda1334$, and Si~\textsc{ii}$\lambda1527$ EWs and their associated errors. 
EW$_{\rm LIS}$ is presented in the top left panel of Figure \ref{fig:lis_his_ebmv_lya} as a function of increasing EW$_{\rm Ly\alpha}$. Here, the relationship between Ly$\alpha$ and EW$_{\rm LIS}$ is presented at $z\sim5$ using the sample from this work, while the $z\sim2\textrm{--}4$ data points are taken from D18. The Ly$\alpha$ EWs uncorrected for IGM absorption are connected by dotted lines, while the intrinsic Ly$\alpha$ EWs calculated by the method described in Section \ref{sec:igm} are connected by solid lines.
It can be seen that at $z\sim5$ the Ly$\alpha$ EW range skews significantly higher than that at lower $z$, with the lowest EW bin at $-1.8\pm1.4$\AA{}. 
The trends at lower redshift are largely unchanged at $z\sim5$. This result implies that the attenuation of Ly$\alpha$ photons (i.e., due to scattering and dust absorption) and the strength of the LIS feature is intrinsically linked, and is fundamental across redshift. In the highest bin of EW$_{\rm Ly\alpha}$, EW$_{\rm Ly\alpha}$ is greater at fixed EW$_{\rm LIS}$ compared to lower redshift, though within 1$\sigma$. 
This deviation suggests intrinsically stronger Ly$\alpha$ at fixed ISM covering fraction \citep[but see e.g.][]{McKinney2019}. We further explore this result in Section \ref{sec:disc_future}.

As seen in Figure \ref{fig:lis_fit}, the shape of the LIS profiles is not strongly dependent on Ly$\alpha$ strength. The ratio of EW$_{\textrm{Si}~\textsc{ii}\lambda1260}$ / EW$_{\textrm{Si}~\textsc{ii}\lambda1527}$ can reflect the changing properties of the gas producing Si~\textsc{ii} absorption: when the ratio approaches $\sim5$, the ratio of their oscillator strengths, Si~\textsc{ii} is in the optically-thin regime. The ratio of EW$_{\textrm{Si}~\textsc{ii}\lambda1260}$ / EW$_{\textrm{Si}~\textsc{ii}\lambda1527}$ in all EW$_{\rm Ly\alpha}$ bins is roughly unity, indicating that these Si~\textsc{ii} features are in the optically-thick regime. This result is consistent with Si~\textsc{ii} line ratios in the $z\sim2-4$ samples presented in D18.

We also examine the evolution of the average spectral properties of the composites after the IGM correction was applied, shown as stars in the top left panel of Figure \ref{fig:lis_his_ebmv_lya}. We see the overall composites following a similar linear relationship at $z\sim2\textrm{--}4$, with the average Ly$\alpha$ emission strength increasing and LIS absorption strength decreasing with increasing redshift. We see an evolution off this linear relationship at $z\sim5$, with the overall composite having stronger Ly$\alpha$ emission and similar LIS absorption properties as the $z\sim4$ sample. We will further examine the average and binned spectral properties in a $z\sim5$ subsample with higher median redshift in Section \ref{sec:disc_future}.

The EW$_{\rm HIS}$ is defined as the weighted average of EW$_{\text{CIV}\lambda1548,1550}$ and EW$_{\text{SiIV}\lambda\lambda1393,1402}$. We display the relationship between EW$_{\rm HIS}$ in bins of Ly$\alpha$ line strength in the lower left panel of Figure \ref{fig:lis_his_ebmv_lya}. In D18, no significant relationship between HIS and Ly$\alpha$ strength was found at $z\sim2\textrm{--}4$. In the $z\sim5$ sample, we find similar strengths of EW$_{\rm HIS}$, showing no redshift evolution in the overall strengths of the features. We also find no trend of stronger EW$_{\rm HIS}$ as a function of stronger EW$_{\rm Ly\alpha}$ emission, similar to what is observed at lower redshift.

In the highest quartile of EW$_{\rm Ly\alpha}$ emission strength, we find evidence for nebular C~\textsc{iv} emission at $z\sim5$. This detection is shown in the bottom panel of Figure \ref{fig:his_fit}. We discuss this spectral feature and its implications for the inferred metallicity of the $z\sim5$ sample in Section \ref{sec:civ}.

\subsection{$E(B-V)$}

Of all the galaxy properties measured by SED fitting, dust attenuation is most directly tied to the makeup of the ISM and CGM. 
As dust grains preferentially scatter bluer light, the amount of reddening reflects the the amount of dust along the line-of-sight. The reddening of the SED of high-redshift sources by dust is typically measured in terms of the quantity $E(B-V)$. The relationship between Ly$\alpha$ and $E(B-V)$ has been well demonstrated: stronger Ly$\alpha$ emission in galaxies correlates with a bluer SED \citep{Shapley2003, Kornei2010, Vanzella2009,Pentericci2007}. The scattering of Ly$\alpha$ photons through neutral gas in the ISM implies greater path lengths to interact with the dust in the galaxy. \citet{Shapley2003} explored the possibility of dust existing in the same phase as the neutral gas itself, implying a direct correlation between the neutral-gas covering fraction and dust reddening for non-resonant radiation, although uncertainty exists between whether the dust lies in outflowing gas or HII regions.

We examine the strength of Ly$\alpha$ in bins of increasing $E(B-V)$ in top right panel of Figure \ref{fig:lis_his_ebmv_lya}. Individual detections of $E(B-V)$ and EW$_{\rm Ly\alpha}$ are overplotted as transparent points. The trends shown by the solid purple line, corresponding to measurements post-IGM correction at $z \sim 5$, are consistent with the  well-established inverse-correlation between EW$_{\rm Ly\alpha}$ and $E(B-V)$ that pertains all the way down to $z\sim0$ \citep{Hayes2014}. The $z\sim2\textrm{--}4$ sample of D18 follows a similar decreasing trend, although the consistency  breaks down at larger $E(B-V)$. Here, EW$_{\rm Ly\alpha}$ is larger at fixed $E(B-V)$ in the $z\sim5$ sample compared to lower redshift. 
We also calculate the correlation between the individual transmitted EW$_{\rm Ly\alpha}$ and $E(B-V)$ measurements in our sample. We perform a Spearman correlation test, with a null hypothesis that EW$_{\rm Ly\alpha}$ and $E(B-V)$ are uncorrelated in the $z\sim5$ sample. The Spearman test produces a correlation coefficient of $-0.27$ with a p-value of 0.00045, strongly rejecting the null hypothesis. This test further supports that the inverse correlation between EW$_{\rm Ly\alpha}$ and $E(B-V)$ continues out to $z\sim5$.

The strength of EW$_{\rm LIS}$ in bins of increasing $E(B-V)$ is presented in the lower right panel of Figure \ref{fig:lis_his_ebmv_lya}. The $z\sim5$ trend also follows the one at $z\sim2\textrm{--}4$, with larger $E(B-V)$ leading to a stronger LIS detection. This result supports the physical picture of dust grains existing in the same phase of gas as that attenuating Ly$\alpha$ and strengthening the LIS absorption \citep{Reddy2016}. This trend also appears to be invariant out to $z\sim5$. The question of which relation is most fundamental --  LIS-$E(B-V)$ or LIS-Ly$\alpha$ -- is difficult to surmise from these plots, but has important implications for the relationship between dust, neutral-phase gas, and escaping Ly$\alpha$ radiation from a galaxy.
In Section \ref{sec:disc_future}, we re-examine the evolution of the relationships among Ly$\alpha$, LIS, and $E(B-V)$ using finer redshift divisions at $z>4$, and gain further insights into this important question.

\subsection{Other Galaxy Properties}

\begin{figure*}
	\centering
	\includegraphics[width=\textwidth]{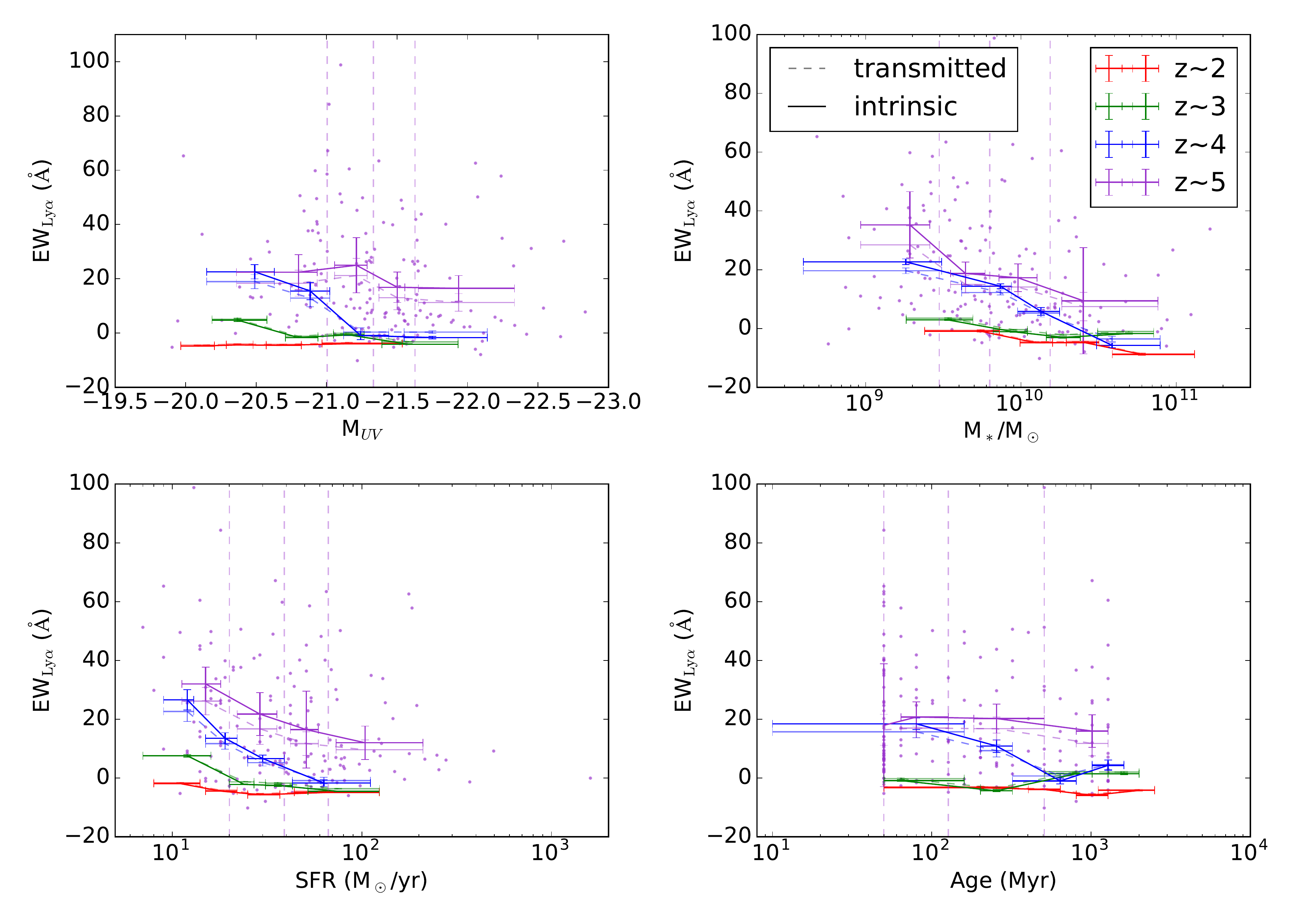}
	\caption{
	EW$_{\rm Ly\alpha}$ as a function of integrated galaxy properties. Within each panel, each redshift sample is divided into four bins of an increasing galaxy property corresponding to that displayed on the x axis: clockwise from top left, M$_{\rm UV}$, M$_*$, age, and SFR. The dashed lines connect measurements of EW$_{\rm Ly\alpha}$ from the composite spectra and the median galaxy property of the objects in each bin. The solid lines represent the same measurements after applying the IGM transmission curves of \citet{Laursen2010}. The purple points are the individual measurements of EW$_{\rm Ly\alpha}$ and galaxy property determined in this work for the $z\sim5$ sample. The vertical, purple, dashed lines represent the borders of the bins of the respective galaxy property. 
	}
	\label{fig:props}
\end{figure*}

In addition to $E(B-V)$, we examine the strength of Ly$\alpha$ binned in terms of galaxy properties of M$_*$, SFR, age, and M$_{\rm UV}$. These relationships out to $z \sim 5$ are presented in Figure \ref{fig:props}. Both the intrinsic and transmitted EW$_{\rm Ly\alpha}$ are shown as solid and dashed lines, respectively, and the individual transmitted measurements for each object are shown as purple points.

Most of these relationships extend those that are present at $z \sim 2\textrm{--}4$. At $z \sim 5$, we find that stronger EW$_{\rm Ly\alpha}$ emission occurs in fainter galaxies, albeit weakly, similar to the trend observed at $z \sim 4$. This trend is skewed towards stronger EW$_{\rm Ly\alpha}$ emission at fixed brightness at $z\sim5$. The relationship between M$_{\rm UV}$ and EW$_{\rm Ly\alpha}$ flattens in the lower redshift samples. We find a  similiar behavior in SFR space, where we find greater EW$_{\rm Ly\alpha}$ emission in galaxies with lower SFR, and greater EW$_{\rm Ly\alpha}$ at fixed SFR at $z \sim 5$ compared to $z \sim 2 - 4$. In stellar-mass space, we find stronger EW$_{\rm Ly\alpha}$ emission in less-massive galaxies, but the trend does not appear to evolve beyond that at $z \sim 4$. Massive galaxies tend to be more dusty \citep{McLure2018,Garn2010,Whitaker2017}, which would suppress their EW$_{\rm Ly\alpha}$.
We find no trend between EW$_{\rm Ly\alpha}$ and age at $z \sim 5$, consistent with results at $z \sim 2\textrm{--}4$. This lack of trend could be explained by EW$_{\rm Ly\alpha}$ reaching equilibrium quickly after the onset of a starburst \citep[50-100~Myr, see][]{Verhamme2008}.

We also examine the correlation between $\Delta v_{\rm Ly\alpha \textrm{--}  LIS}$ and EW$_{\rm Ly\alpha}$ and recover the anti-correlation of \citet{Shapley2003}, indicating that a Ly$\alpha$ velocity peak closer to systemic indicates an easier pathway of escape for Ly$\alpha$ photons. This anti-correlation is weak, with a Spearman correlation test producing a correlation coefficient of -0.17 and a p-value of 0.14.

In summary, at $z \sim 5$, we find similar trends to those at $z \sim 4$: galaxies with stronger Ly$\alpha$ emission tend to have lower SFRs, less star formation, bluer UV continua, and less dust extinction. We also find stronger EW$_{\rm Ly\alpha}$ emission at fixed galaxy property in the $z\sim5$ sample.

\section{Discussion} \label{sec:disc_future}

Cosmic reionization is the phase change of neutral to ionized hydrogen in the IGM, which appears to be roughly finished by $z\sim6$ \citep{Fan2006,McGreer2015}. The leading theory is that reionization was caused by radiation from massive stars at $\lambda\leq912$\AA{} escaping from galaxies and ionizing the surrounding IGM \citep{Finkelstein2012,Finkelstein2019,Robertson2015a}. 
Due to the high opacity of the IGM at $z>6$ to photons just below the Lyman limit, it is difficult to directly observe the fraction of leaking ionizing to non-ionizing radiation in high-redshift galaxies \citep{Vanzella2012}. 
However, the Ly$\alpha$ feature is well detected out to redshifts near the end of the epoch of reionization \citep[i.e., $z\sim6$;][]{Stark2010a,Schenker2014,DeBarros2017} and while significantly more challenging, even into the epoch of reionization \citep[e.g.,][]{Jung2018}.
By modeling the rest-UV spectra of LBGs to infer the intrinsic ionizing luminosity, \citet{Steidel2018} has shown that the fraction of escaping ionizing radiation in a galaxy at $z\sim3$ is an increasing function of EW$_{\rm Ly\alpha}$. This result supports a physical picture of both Ly$\alpha$ photons and ionizing radiation escaping through ``holes" in the neutral ISM, similar to that discussed in \citet{Reddy2016}. This picture is empirically supported the Ly$\alpha$ profile and Lyman continuum emission of Ion2 in \citet{Vanzella2019}.
Thus, understanding the intrinsic Ly$\alpha$ strength of high-redshift galaxies can provide important clues to the evolution of leaking ionizing radiation, and consequently the epoch of reionization \citep{Jung2018,Mason2018,Laursen2019}. 

\subsection{Splitting $z\sim5$ sample into low- and high-redshift} \label{sec:split}

Given the size of the $z\sim5$ sample (including over twice as many galaxies as in the $z\sim4$ sample from \citet{Jones2012}) and our desire to obtain finer redshift sampling with a high-redshift bin that approaches the epoch of reionization, we isolate high-redshift subsamples within our $z\sim5$ dataset.
We divide the $z\sim5$ sample at its median redshift. Our ``low-redshift" $z\sim5$ bin contains 87 galaxies with $z_{\rm med}=4.34$, and our ``high-redshift" $z\sim5$ bin contains 88 galaxies with $z_{\rm med}=4.73$.

\begin{figure*}
	\centering
	\includegraphics[width=\textwidth]{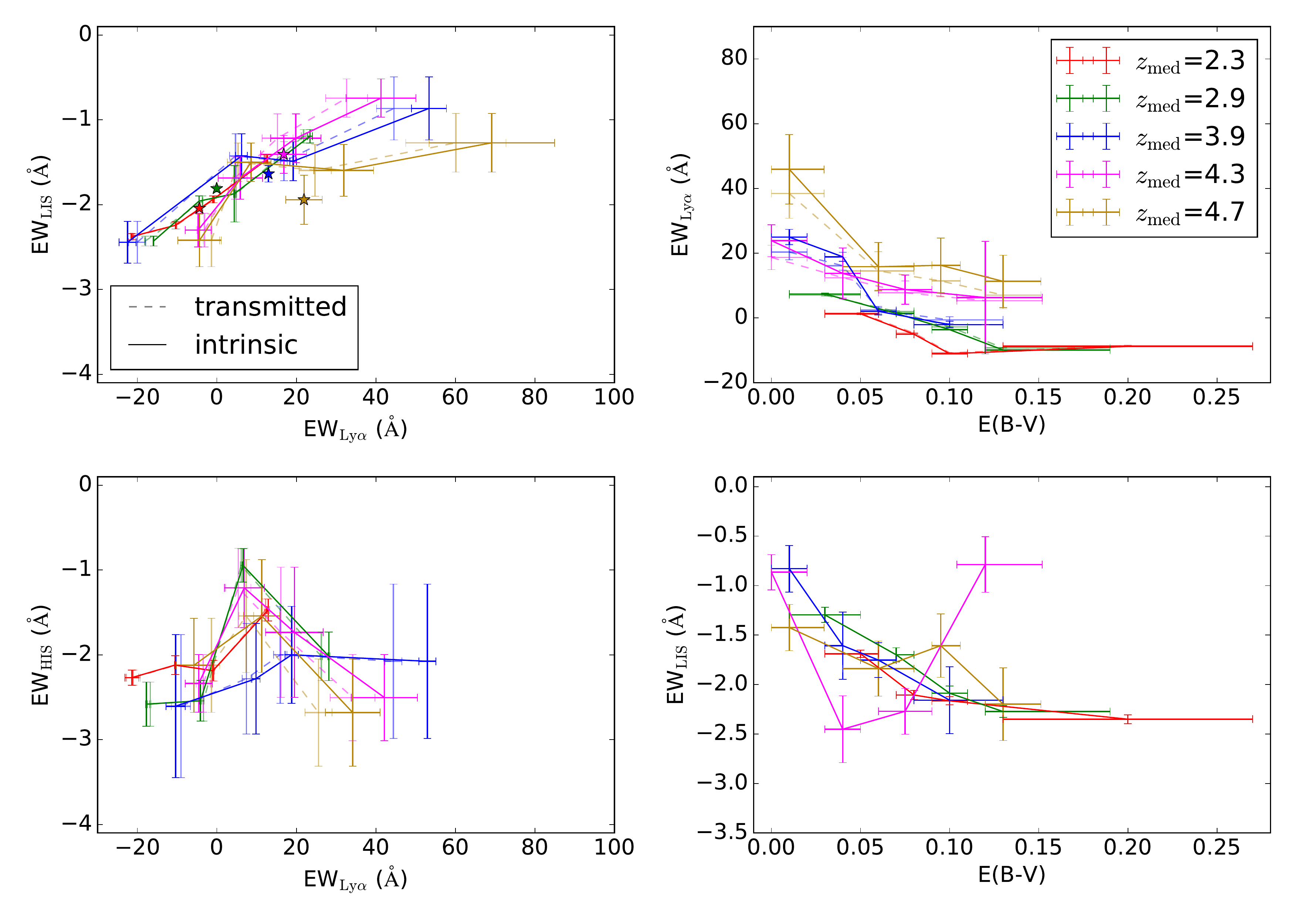}
	\caption{
		Relations between EW$_{\rm Ly\alpha}$, EW$_{\rm LIS}$, EW$_{\rm HIS}$, and $E(B-V)$ with the two redshift subsamples at $z\sim5$. The magenta and gold points are the $z\sim5$ sample divided into a lower and higher redshift sample of $z_{\rm med}=4.34$ and 4.73. EW$_{\rm Ly\alpha}$, EW$_{\rm LIS}$, and EW$_{\rm HIS}$ were measured from composite spectra. The measurements from the observed composites are displayed in transparent points connected by dotted lines, while the corrected measurements after application of degraded IGM transmission curves are the opaque points connected by solid lines. The red, green, and blue points are measurements of the $z\sim2$, 3, and 4 samples of D18. 
		\textbf{Top left panel:} EW$_{\rm LIS}$ vs. EW$_{\rm Ly\alpha}$ with two redshift samples of $z\sim5$.  Each redshift sample was divided into four bins of increasing EW$_{\rm Ly\alpha}$. Both EW$_{\rm Ly\alpha}$ and EW$_{\rm LIS}$ are measured from composite spectra made from combining the individual spectra of the objects in each EW$_{\rm Ly\alpha}$ bin. Also included are average measurements of the composites of the $z\sim2\textrm{--}4$ samples and the $z_{\rm med}=4.34$ and 4.73 subsamples, displayed as stars. 
		\textbf{Bottom left panel:} EW$_{\rm HIS}$ vs. EW$_{\rm Ly\alpha}$. The binned samples are the same as the top panel, with the exception that the objects were required to have coverage of C~\textsc{iv}$\lambda1548,1550$. EW$_{\rm HIS}$ and EW$_{\rm Ly\alpha}$ are also measured from composite spectra, made from combining the individual spectra of the objects in each EW$_{\rm Ly\alpha}$ bin. The EW$_{\rm HIS}$-EW$_{\rm Ly\alpha}$ measurement in the highest EW$_{\rm Ly\alpha}$ bin of the $z_{\rm med}=4.73$ subsample was removed due to contamination from CIV emission.
		\textbf{Top right panel:} EW$_{\rm Ly\alpha}$ vs. $E(B-V)$. Each sample was divided into four bins of increasing $E(B-V)$, with EW$_{\rm Ly\alpha}$ measured from composite spectra and $E(B-V)$ from the median of the objects in each bin. The transparent points are individual detections.
		\textbf{Bottom right panel:} EW$_{\rm LIS}$ vs. $E(B-V)$. The binned samples are the same as the top panel, with EW$_{\rm LIS}$ also measured from composite spectra.
	}
	\label{fig:lis_his_ebmv_lya_2z}
\end{figure*}

The strengths of interstellar metal lines and Ly$\alpha$ in the lower- and higher-redshift $z\sim5$ subsamples are presented in the left panels of Figure \ref{fig:lis_his_ebmv_lya_2z}. The purple $z\sim5$ line of Figure \ref{fig:lis_his_ebmv_lya} at  $z_{\rm med}=4.52$ is replaced by a magenta curve for the $z_{\rm med}=4.34$ sample and a gold curve for the $z_{\rm med}=4.73$ sample.
At $z_{\rm med}\leq4.34$, the positive correlation between EW$_{\rm LIS}$ and EW$_{\rm Ly\alpha}$ is preserved. This unchanging correlation reflects the fundamental connection between the amount of neutral gas and the attenuation of the Ly$\alpha$ line. In the $z_{\rm med}=4.73$ sample, the trend begins to evolve at EW$_{\rm Ly\alpha}\geq20$\AA{}. We observe stronger Ly$\alpha$ emission at fixed LIS absorption strength.

The average properties of the composite spectra out to $z\sim5$ are displayed as stars in the upper left panel of Figure \ref{fig:lis_his_ebmv_lya_2z}. At $z_{\rm med}\leq4.34$, the average Ly$\alpha$ emission strength increases and the average LIS absorption strength decreases with increasing redshift. 
This trend indicates that the neutral-gas covering fraction is decreasing with increasing redshift. This relationship appears to be different at $z_{\rm med}=4.73$, where the average properties of the $z_{\rm med}=4.73$ sample are offset from the trend of decreasing EW$_{\rm Ly\alpha}$ with increasing EW$_{\rm LIS}$ established at $z_{\rm med}\leq4.34$. This result implies additional factors beyond neutral-gas covering fraction are influencing the average EW$_{\rm Ly\alpha}$ and EW$_{\rm LIS}$ of the $z_{\rm med}=4.73$ subsample.

We further probe the state of the ISM near the epoch of reionization based on the mutual relationships among EW$_{\rm Ly\alpha}$, EW$_{\rm LIS}$, and $E(B-V)$. The upper right panel of Figure \ref{fig:lis_his_ebmv_lya_2z} shows stronger  EW$_{\rm Ly\alpha}$ emission in bluer galaxies, similar to that of Figure \ref{fig:lis_his_ebmv_lya}, with the two $z\sim5$ subsamples more clearly illustrating the evolution at $z_{\rm med}\geq4.34$. Galaxies at $z_{\rm med}\geq4.34$ have stronger Ly$\alpha$ emission at fixed $E(B-V)$. This evolution is not present in the relationship between EW$_{\rm LIS}$ and $E(B-V)$ shown in the lower right panel of Figure \ref{fig:lis_his_ebmv_lya_2z}: stronger LIS absorption correlates with more reddening, invariant with redshift.

\begin{figure*}
	\centering
	\includegraphics[width=\textwidth]{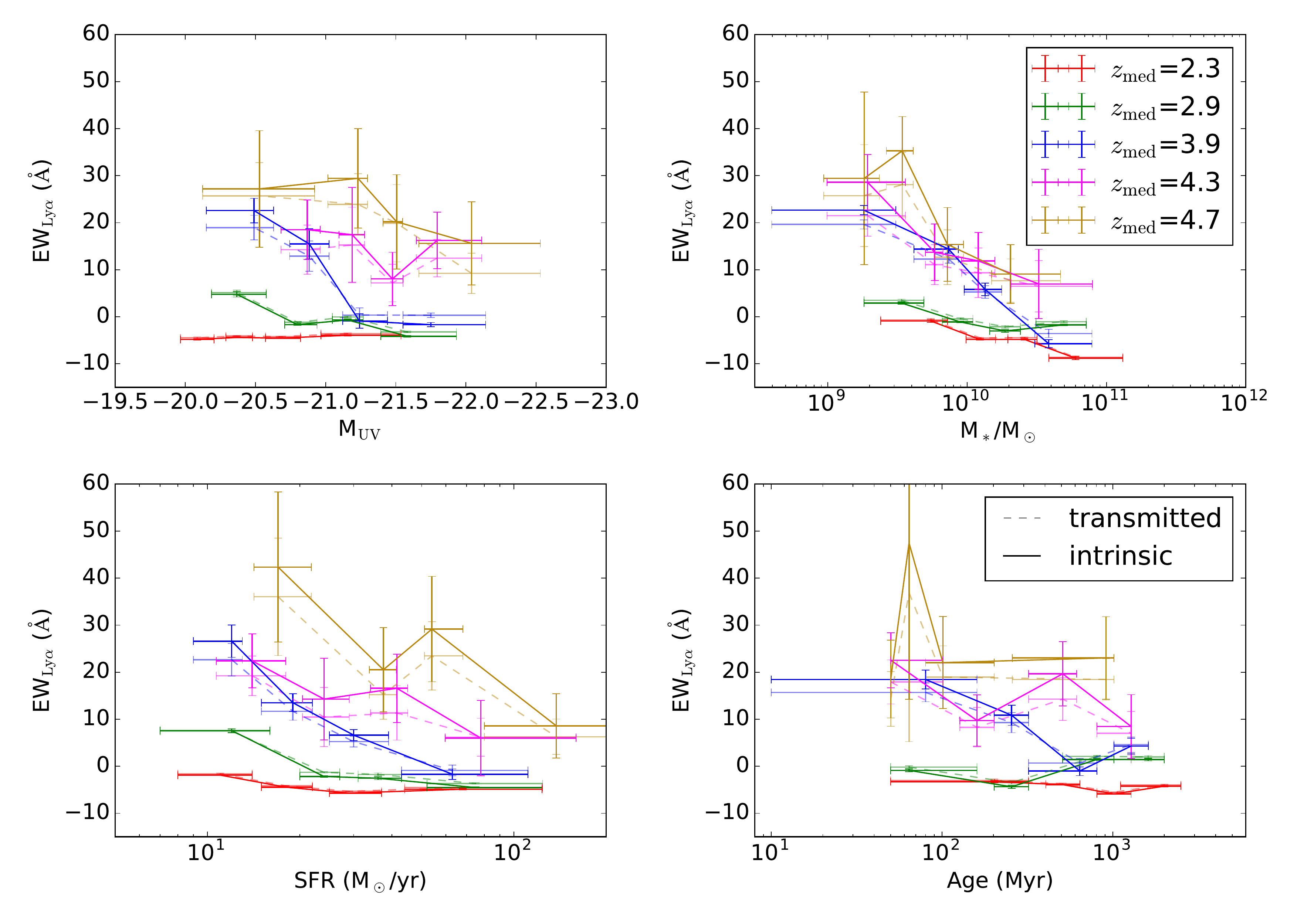}
	\caption{
		EW$_{\rm Ly\alpha}$ as a function of integrated galaxy properties. The magenta and gold points correspond to the purple points of Figure \ref{fig:props} split into two redshift samples of $z_{\rm med}=4.34$ and 4.73. Within each panel, each redshift sample is divided into four bins of an increasing galaxy property corresponding to that displayed on the x axis: clockwise from top left, M$_{\rm UV}$, M$_*$, age, and SFR. The dashed lines connect measurements of EW$_{\rm Ly\alpha}$ from the composite spectra and the median galaxy property of the objects in each bin. The solid lines represent the same measurements after applying the IGM transmission curves of \citet{Laursen2010}.
	}
	\label{fig:props_2z}
\end{figure*}

The redshift evolution of EW$_{\rm Ly\alpha}$ and other galaxy properties of Figure \ref{fig:props} also sheds light on the physical differences of galaxies as we look back towards the epoch of reionization. The trends found in the single $z\sim5$ sample analysis are more strongly demonstrated in Figure \ref{fig:props_2z}, based on the lower- and higher-redshift $z\sim5$ subsamples. The pattern of lower EW$_{\rm Ly\alpha}$ in brighter, more massive, highly star-forming galaxies is preserved across all redshift samples. This trend can also be seen down to $z\sim0$ \citep{Hayes2014}. Additionally, the evolution towards greater EW$_{\rm Ly\alpha}$ at fixed M$_{\rm UV}$, SFR, and age with increasing redshift is shown more clearly with the $z_{\rm med}=4.34$ and $z_{\rm med}=4.73$ subsamples. 
While we don't see a correlation of increased EW$_{\rm Ly\alpha}$ with increased specific star-formation rate (sSFR), we note that EW$_{\rm Ly\alpha}$ is elevated in the $z_{\rm med}=4.73$ subsample compared to the $z_{\rm med}=4.34$ subsample at fixed sSFR, indicating that there are other galaxy properties affecting the strength of Ly$\alpha$.
On the other hand, the relationship between EW$_{\rm Ly\alpha}$ and M$_{*}$ does not evolve in the three highest redshift subsamples. The trends of EW$_{\rm Ly\alpha}$ and M$_{\rm UV}$, SFR, sSFR, and age reflect that EW$_{\rm Ly\alpha}$ is increasing at fixed galaxy property, especially at $z_{\rm med}\geq4.34$. This is similar to the rest-UV spectral trends of Figure \ref{fig:lis_his_ebmv_lya_2z}, where it is shown that EW$_{\rm Ly\alpha}$ is increasing at fixed ISM/CGM property.

We also detect nebular C~\textsc{iv} emission in the highest EW$_{\rm Ly\alpha}$ quartile of the $z_{\rm med}=4.73$ subsample. This profile, shown in Figure \ref{fig:civ_highz}, is similar to that found within the total $z\sim5$ sample, shown in  Figure \ref{fig:his_fit}. We discuss the implications of these detections for the inferred metallicity of the $z\sim5$ sample in Section \ref{sec:civ}.

\subsection{Intrinsic Ly$\alpha$ Production Evolution} \label{sec:lya_intrins}

The three factors that affect the observed strength of Ly$\alpha$ in a galaxy spectrum are (1) the intrinsic production rate of Ly$\alpha$ photons in HII regions, which is set by the efficiency of ionizing photon production for a given SFR, (2) the radiative transfer of Ly$\alpha$ photons through the ISM and CGM of the galaxy, and (3) the transfer of Ly$\alpha$ photons through the IGM. We have corrected for IGM opacity through application of the theoretical transmission curves of \citet{Laursen2010}, which, calculated as an average of many sight lines through the IGM, approximate the effects of the IGM on our stacked spectra composed of many galaxies.
The anti-correlation between EW$_{\rm Ly\alpha}$ and EW$_{\rm LIS}$, which remains roughly constant from $z_{\rm med}=2.3$ to $z_{\rm med}=4.3$, can be seen as consequence of ISM and CGM radiative transfer. An increased covering fraction of neutral ISM and CGM gas both strengthens LIS absorption and weakens Ly$\alpha$ emission. Based on the lack of evolution in the relationship between EW$_{\rm Ly\alpha}$ and EW$_{\rm LIS}$, we argue that this correlation is driven by variations in covering fraction of interstellar and circumgalactic gas, and that the intrinsic typical Ly$\alpha$ production efficiency of samples at these redshifts is similar. This is supported by the distinction in \citet{Trainor2019} between quantities related to Ly$\alpha$ {\it production} and {\it escape} that jointly modulate observed Ly$\alpha$ strength. The authors argue that EW$_{\rm LIS}$ is strictly an escape-related quantity, supporting our conclusion that changing ISM/CGM properties should entirely define the EW$_{\rm Ly\alpha}$ vs. EW$_{\rm LIS}$ relationship assuming intrinsic Ly$\alpha$ production is constant.

The invariance of the EW$_{\rm Ly\alpha}$ vs. EW$_{\rm LIS}$ relation breaks down in the highest-redshift bin at $z_{\rm med}=4.73$. In the highest bin of EW$_{\rm Ly\alpha}$, the $z_{\rm med}=4.73$ sample differs significantly from the relationship defined at lower redshift.
Additionally, at $z_{\rm med}=4.73$, the average EW$_{\rm Ly\alpha}$ and EW$_{\rm LIS}$ deviate strongly from the linear relationship between EW$_{\rm Ly\alpha}$ and EW$_{\rm LIS}$ defined by the lower redshift samples. These deviations, which are at $\geq\sim1\sigma$, suggest the presence of additional factors modulating the strength of Ly$\alpha$ emission, along with radiative transfer through the ISM, CGM, and IGM. Specifically, this difference suggests a significant increase in the intrinsic efficiency of Ly$\alpha$ photon production, and, correspondingly, ionizing radiation.

A scenario of increased intrinsic Ly$\alpha$ production efficiency among the strongest Ly$\alpha$ emitters at $z_{\rm med}=4.73$ is additionally favored by the joint consideration of the EW$_{\rm Ly\alpha}$ vs. EW$_{\rm LIS}$ and EW$_{\rm Ly\alpha}$ vs. $E(B-V)$ relations. The dust and neutral-gas content in the ISM of a galaxy is intimately connected, with both thought to exist in the same physical location in the galaxy. The relationship between EW$_{\rm LIS}$ and $E(B-V)$ is thus expected to be invariant across redshift, as it is entirely defined by the neutral-gas covering fraction (and consequently the amount of neutral-gas and dust) of the galaxy. The relationship in the bottom right panel of Figure \ref{fig:lis_his_ebmv_lya_2z} shows exactly this pattern. 
On the other hand, the redshift evolution towards stronger Ly$\alpha$ emission at fixed $E(B-V)$ shows that, in addition to the dust and neutral-gas content of the galaxy modulating the strength of Ly$\alpha$, the intrinsic production rate also increases across redshift (assuming all else being equal). While our results suggest that the dust and neutral-phase gas exist co-spatially, the dust may be present in many locations throughout the ISM/CGM. Fully disentangling the interplay between EW$_{\rm Ly\alpha}$, EW$_{\rm LIS}$, and $E(B-V)$ will require further work (i.e., individual detections of EW$_{\rm LIS}$).

The observed difference in intrinsic Ly$\alpha$ photon production rate has important implications for understanding the contribution of galaxies to the reionization of the Universe. The comoving ionizing emissivity of the Universe can be modeled as the cosmic SFR ($\rho_{\rm SFR}$) multiplied by the number of LyC photons produced per unit SFR ($\xi_{\rm ion}$) multiplied by the escape fraction of LyC photons from the galaxy ($f_{\rm esc}$). Physically, the EW of Ly$\alpha$ describes the amount of escaping Ly$\alpha$ emission relative to the non-ionizing UV continuum (analogous to $\rho_{\rm SFR}$ when averaged). 
The production of these Ly$\alpha$ photons in HII regions comes from recombination of protons and electrons, driven by LyC photons ionizing the hydrogen gas.
Given that $f_{\rm esc}$ is directly connected to the properties of the ISM and CGM demonstrated by the relationships of EW$_{\rm Ly\alpha}$,  EW$_{\rm LIS}$, and $E(B-V)$ at low redshift, an evolving EW$_{\rm Ly\alpha}$ at fixed EW$_{\rm LIS}$ and $E(B-V)$ indirectly measures the evolving $\xi_{\rm ion}$ with redshift. 
The evolution of $\xi_{\rm ion}$ is an important input into models of reionization \citep{Bouwens2015a,Stanway2016}, and should not be taken as constant \citep[e.g.,][]{Robertson2015a}.

\subsection{Metallicity of the $z\sim5$ Sample} \label{sec:civ}

The relationship between UV slope and UV luminosity evolves towards bluer colors at fixed luminosity towards higher redshift \citep{Bouwens2014}. The observed evolution towards decreasing median $E(B-V)$ from $z\sim2$ to $z\sim5$ in this work is consistent with these results.
This evolution could reflect a fundamental change in dust and metal content of these galaxies, or could reflect a bluer intrinsic stellar population at higher redshift. 
Spectroscopic observations are required to distinguish among different scenarios of dust, metallicity, or stellar population evolution.

\begin{figure}
	\centering
	\includegraphics[width=\columnwidth]{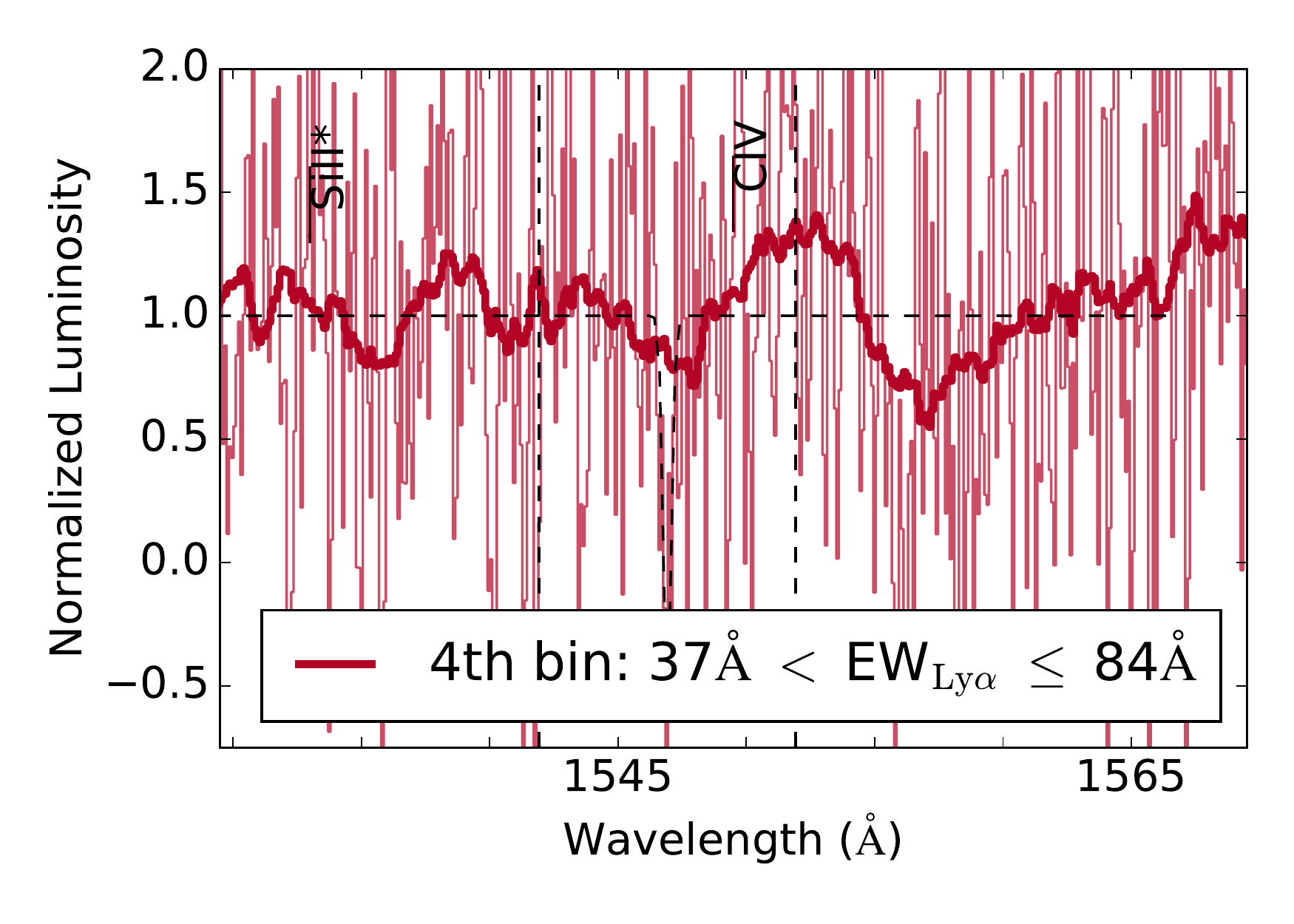}
	\caption{
		Nebular C~\textsc{iv} emission bluewards of the interstellar C~\textsc{iv} absorption feature of the $z_{\rm med}=4.73$ subsample. Displayed is composite of the objects in the highest EW$_{\rm Ly\alpha}$ quartile. As in Figure \ref{fig:his_fit}, the thin curve represents the continuum-normalized, rest-frame composite spectrum. The opaque spectrum is convolved with a boxcar window of width 3\AA{} for easier viewing. The vertical, dashed lines  illustrate the $\pm5$\AA{} integration windows from which EW measurements would have been made, but due to contamination from the nebular C~\textsc{iv} emission, the interstellar C~\textsc{iv} measurement was removed from the analysis.
	}
	\label{fig:civ_highz}
\end{figure}

The detection of nebular C~\textsc{iv} emission in the highest bin of EW$_{\rm Ly\alpha}$ at $z\sim5$ provides independent evidence for significantly sub-solar metallicity.
Nebular C~\textsc{iv} emission is produced by extreme radiation fields in star-forming galaxies \citep{Berg2019} and detections at lower redshift are typically found in metal-poor galaxies  \citep{Erb2010,Stark2014, Senchyna2017, Senchyna2019, Berg2019, Vanzella2016, Vanzella2017}. \citet{Stark2015} detected nebular C~\textsc{iv} in one galaxy at $z=7.045$, out of a sample of four $z\sim6-8$ galaxies, whereas the typical C~\textsc{iv} detection rate in UV-selected galaxies at $z\sim2\textrm{--}3$ is only 1 per cent \citep{Steidel2002, Hainline2011}. A single, strong C~\textsc{iv} detection in a sample of four $z\sim6-8$ galaxies could imply a larger population of lower-metallicity, nebular C~\textsc{iv}-emitting galaxies during the epoch of reionization. 
In our overall $z_{\rm med}=4.52$ sample, we see strong nebular C~\textsc{iv} emission in the highest EW$_{\rm Ly\alpha}$ bin (near rest-frame 1549\AA{} in the bottom panel of Figure \ref{fig:his_fit}). In the $z_{\rm med}=4.73$ subsample, this feature is also prominent, as shown in Figure \ref{fig:civ_highz}. A significant detection in a bin containing 25 per cent of the objects in the $z\sim5$ sample indicates a changing population of galaxies at $z\sim5$ with greater nebular C~\textsc{iv}, harder ionizing fields, and lower metallicity. 
When compared to the highest EW$_{\rm Ly\alpha}$ quartile of \citet{Steidel2002} at $z\sim2$, we find an elevated nebular C~\textsc{iv} to Ly$\alpha$ flux ratio of $4.0\pm2.4$ per cent and $2.3\pm1.7$ per cent in the  $z_{\rm med}=4.52$ and $z_{\rm med}=4.73$ samples respectively, compared to $\leq1$ per cent at $z\sim2$. The EW of this nebular C~\textsc{iv} line was measured using the method in Section \ref{sec:absew} as $1.98\pm1.13$\AA{} and $1.61\pm1.08$\AA{} in the $z_{\rm med}=4.52$ and $z_{\rm med}=4.73$ samples, respectively.

Systematically lower metallicity in galaxies near the epoch of reionization has important implications for $\xi_{\rm ion}$. The connection between metallicity and ionizing photon production is more easily studied at lower redshift, where rest-optical metal lines can be observed with ground-based optical telescopes. \citet{Shivaei2018} measured photon production efficiencies in star-forming galaxies at $z\sim2$ and found galaxies with elevated $\xi_{\rm ion}$ have bluer UV spectral slopes, higher ionization parameters, and lower metallicities. Extreme galaxies at lower redshift with subsolar metallicity have also been demonstrated to have higher average EW$_{\rm Ly\alpha}$ \citep{Erb2010}.
These observations at lower redshift paint a physical picture of galaxies with lower metallicity producing an intrinsically harder ionizing spectra per unit star-formation rate,  emitting Ly$\alpha$ photons that are also less likely to be absorbed during multiple scatterings in less-dusty ISM/CGM. This picture is once again supported by the framing of \citet{Trainor2019}, according to which the evolution of the average EW$_{\rm Ly\alpha}$ vs. EW$_{\rm LIS}$ properties across redshift reveals the changing {\it escape}-related properties of the ISM/CGM, while the evolving metallicity supports changing {\it production}-related parameters of the stellar population at high redshift. If these lower-metallicity, lower neutral-gas covering-fraction galaxies are more common at $z\sim5$, as suggested by the evolving $E(B-V)$ and nebular C~\textsc{iv} emission in our sample, they will contribute more strongly to reionization because of increased $\xi_{\rm ion}$ and $f_{\rm esc}$.
Direct metallicity measurements at $z\geq5$ are vital to disentangle and quantify these effects.

\section{Summary} \label{sec:summary}

The rest-UV spectra of star-forming galaxies contain important information on star formation, dust and metal content of the ISM and CGM, and gas kinematics. Interpreting this information provides clues into the changing galaxy properties closer to the epoch of reionization. These properties include those related to the escape of ionizing radiation, such as neutral-gas covering fraction, and properties related to production of ionizing radiation, such as intrinsic Ly$\alpha$ production rate. To study these properties, we analyzed the relative strengths of Ly$\alpha$, LIS, and HIS lines in a sample of $z\sim5$ star-forming galaxies for the first time and compared them to similar analyses at $z\sim2\textrm{--}4$ performed in D18. We also performed corrections for IGM attenuation on the Ly$\alpha$ profiles of sample composites using transmission curves of \citet{Laursen2010}. The key results are as follows:
\begin{enumerate}
	\item We find that the redshift invariance of the relationship of EW$_{\rm LIS}$ vs. EW$_{\rm Ly\alpha}$ deviates at $z_{\rm med}=4.52$, and even more significantly at $z_{\rm med}=4.73$, implying an evolving intrinsic Ly$\alpha$ photon production rate at $z\sim5$ (assuming all else being equal).
	The EW$_{\rm LIS}$ vs. EW$_{\rm Ly\alpha}$ relation illustrates the effects of the neutral-gas covering fraction on the rest-UV spectrum, and greater EW$_{\rm Ly\alpha}$ at fixed EW$_{\rm LIS}$ at $z\sim5$ implies a greater intrinsic Ly$\alpha$ production rate. 
	\item We find that the relationship of EW$_{\rm Ly\alpha}$ vs. $E(B-V)$ evolves at higher redshift, with increased EW$_{\rm Ly\alpha}$ at fixed $E(B-V)$ at $z\sim5$, but no redshift evolution in the relationship of EW$_{\rm LIS}$ vs. $E(B-V)$, suggesting greater intrinsic Ly$\alpha$ production at fixed ISM/CGM properties using the non-evolution of EW$_{\rm LIS}$ and EW$_{\rm Ly\alpha}$ as a control.
	The invariance of EW$_{\rm LIS}$ vs. $E(B-V)$ reflects the fundamental connection between neutral-phase gas content and dust in the ISM and CGM of star-forming galaxies at high redshift. However, we find greater EW$_{\rm Ly\alpha}$ at fixed $E(B-V)$ at $z\sim5$, implying for a given amount of dust and neutral gas, there are more Ly$\alpha$ photons being produced.
	\item The robust detection of nebular C~\textsc{iv} in the highest quartile of EW$_{\rm Ly\alpha}$ objects within the $z_{\rm med}=4.52$ and the higher-redshift $z_{\rm med}=4.73$ subsample supports a scenario of significantly subsolar metallicity in $\sim25$ per cent of our sample of star-forming galaxies near the epoch of reionization. Among non-AGNs, nebular C~\textsc{iv} is typically found in a lower-metallicity population in galaxies from $z\sim0-2$. Additionally, we find evolving UV slopes seen through the changing median $E(B-V)$ from z$\sim2-5$, also consistent with a more metal-poor population of galaxies at higher-redshift. 
	\item The relationship of EW$_{\rm HIS}$ vs. EW$_{\rm Ly\alpha}$ remains invariant with redshift out to $z\sim5$, implying that ionized-phase and neutral-phase gas in the ISM and CGM remains in physically-distinct regions out to higher redshift.
	\item Higher EW$_{\rm Ly\alpha}$ is correlated with fainter M$_{\rm UV}$, lower M$_{*}$, lower SFR, and younger age at $z\sim5$, similar to the trends found at $z\sim4$. We also see an evolution of greater EW$_{\rm Ly\alpha}$ at fixed SFR, M$_{\rm UV}$, and age at $z\sim5$. This evolution of greater EW$_{\rm Ly\alpha}$ at fixed galaxy property is consistent with the conclusion that the intrinsic Ly$\alpha$ production rate is increasing at $z\sim5$.
\end{enumerate}

These results have important implications for the understanding of the evolving galaxy population just after the completion of reionization. Higher intrinsic Ly$\alpha$ photon production rate implies a greater $\xi_{\rm ion}$ at $z\sim5$, a vital input parameter to models of reionization.
The analyses performed here are especially revealing $in$ the epoch of reionization, in which the star-forming galaxies are thought to be the most significant contributers to the comoving ionizing emissivity of the universe. The upcoming launch of the {\it James Webb Space Telescope}, with its groundbreaking near-IR spectroscopic capabilities, will allow for the measurement of rest-UV spectral features out to $z\sim10$, and we will discover whether the trends at $z\sim5$ continue into the epoch of reionziation.
\linebreak

AES, AJP, XD, and MWT acknowledge support for program GO-15287, provided by NASA through a grant from the Space Telescope Science Institute (STScI), which is operated by the Association of Universities for Research in Astronomy, Inc., under NASA contract NAS 5-26555.
We also acknowledge a NASA contract supporting the ``WFIRST Extragalactic Potential Observations (EXPO) Science Investigation Team" (15-WFIRST15-0004), administered by GSFC. We wish to extend special thanks to those of Hawaiian ancestry on
whose sacred mountain we are privileged to be guests. Without their generous hospitality, most
of the observations presented herein would not have been possible.


\end{document}